\documentstyle[preprint,eqsecnum,prb,aps]{revtex}
\begin{document}
\preprint{IUCM94-023}
\draft
\title{Impurities in $s=1$ Heisenberg Antiferromagnets}
\author{Erik S.\ S\o rensen}
\address{Department~of~Physics,
Indiana~University, Bloomington, IN~47405}
\author{Ian~Affleck}
\address{Department of Physics and Canadian Institute
for Advanced Research}
\address{
University of British Columbia, Vancouver, BC, V6T 1Z1, Canada}
\date{December 20, 1994}
\maketitle
\begin{abstract}
The $s=1$ Heisenberg Antiferromagnet is studied in the presence
of two kinds of local impurities.
{}First, a perturbed antiferromagnetic bond with $J'\ne J$ at the center
of an even-length open chain is considered.
Using the density
matrix renormalization group method we find that, for sufficiently strong
or weak $J'$, a bound state is
localized at the impurity site, giving rise to an energy level in
the Haldane gap. The energy of the bound state
is in agreement with perturbative results, based on $s=1/2$ chain-end
excitations, both in the weak and strong coupling limit.
In a region around the uniform limit, $J'=J$, no states are found
with energy below the Haldane gap.
Secondly, a $s=1/2$ impurity at the center of an
otherwise even-length open
chain is considered. The coupling to the $s=1/2$ impurity is varied.
Bound states in the Haldane gap are found {\it only} for sufficiently weak
(antiferromagnetic) coupling.
{}For a $s=1/2$ impurity coupled with a strong (antiferromagnetic)
bond, {\it no} states are found in the Haldane.
Our results are in good qualitative agreement with
recent experiments on doped NENP and Y$_2$BaNiO$_5$.
\end{abstract}

\section{Introduction}
In a seminal paper, Haldane showed that integer spin chains
display a gap in the spectrum, where as half-integer spin chains are
gapless~\cite{haldane}.
This was confirmed both by experiments~\cite{expgap,renard87}
and numerical calculations~\cite{numgap}. Currently,
the best estimates of
the gap for the isotropic $s=1$ chain  are
$\Delta=0.41050(2)J$, using the density matrix renormalization
group (DMRG) method~\cite{whitehuse}, and $\Delta=0.41049(2)J$ using
exact diagonalization~\cite{golinelli1}. The dynamical structure
factor has also been determined in great detail, by numerical studies of
finite systems~\cite{skw}.
Most of the experimental effort has in recent years
focused on Ni(C$_2$H$_8$N$_2$)$_2$NO$_2$(ClO$_4$) (NENP)
which show clear evidence of the Haldane gap~\cite{renard87,expNENP}.
NENP has not been observed to undergo a three-dimensional
ordering at any accessible temperature as opposed to CsNiCl$_3$ which
yielded the first experimental confirmation of the Haldane gap~\cite{expgap}.
NENP has a fairly large single ion anisotropy estimated
to be $D/J=0.18$~\cite{golinelli2} where as CsNiCl$_3$ is almost isotropic.
Recently, experiments have been performed on
Y$_2$BaNiO$_5$~\cite{spuche1,spuche2,buttrey,darriet,batlogg,ditusa1,ditusa}.
This charge-transfer insulator has long one-dimensional chains of
Ni atoms at the center of compressed corner sharing oxygen octahedra.
Thus, along the chains, the Ni$^{2+}$ ions, with $s=1$, are separated by
nonmagnetic O$^{2-}$ ions.
The antiferromagnetic (AF) superexchange interaction
between the Ni$^{2+}$ ions is then mediated through the oxygen sites.
Y$_2$BaNiO$_5$ has a fairly large exchange coupling of
$J/k_b$=285K~\cite{darriet}-322K~\cite{spuche2}, estimated from
high temperature susceptibility measurements.
{}From inelastic neutron scattering
experiments on powders~\cite{darriet}
it is known that the inter chain coupling is small, $J'/J\ll10^{-2}$.
Thus, the compound is highly one-dimensional and no three-dimensional
ordering has been seen.
In powder averaged inelastic neutron scattering
experiments~\cite{darriet} two gaps have been observed;
{}For fluctuations parallel to the
chain, $\Delta_{||}=186\pm 12K$, and for fluctuations perpendicular to
the chain, $\Delta_{\perp}=99\pm 6K$. This is in good agreement
with single crystal measurements of $\Delta_{\perp}=100\pm5K$~\cite{batlogg}.
The two gaps are explained in terms of a single ion anisotropy
$D/J=0.16\pm0.02$~\cite{darriet},
a value close to what is found for NENP.

The one-dimensional Haldane gap systems are by now fairly well
understood. However, the important effect of impurities on these systems
have only recently been considered.
Y$_2$BaNiO$_5$ offers several possibilities for doping
and is thus well suited for such studies.
Non-magnetic Zn$^{2+}$ ions can be substituted for Ni$^{2+}$,
thus effectively severing the Ni chains.
Experiments have been performed on Zn doped samples,
Y$_2$BaNi$_{1-x}$Zn$_x$O$_5$~\cite{spuche2,batlogg,ditusa1,ditusa,ramirez},
confirming the picture of chain cutting, with a possibility for
an  {\it increase} in the gap with doping~\cite{ditusa1} due to the
fact that the average length of the chain segments is small enough
for finite-size effects to be observable.
Severing of the chains is predicted to have rather dramatic effects
on the low energy excitations. It can be shown that
the AF Heisenberg model is fairly well approximated by
a valence bond solid (VBS) model~\cite{aklt}.
The breaking of the valence bonds by the introduction
of impurities should produce free $s=1/2$ spins at the sites
next to the impurity. The effect of such free $s=1/2$ at the
chain-ends has been observed in electron-spin-resonance
(ESR) experiments on NENP by doping with Cu~\cite{hagiwara}, thereby
introducing a $s=1/2$ at the impurity site, and by doping with non-magnetic
Zn, Cd and Hg~\cite{glarum}. These findings have recently been contrasted
by specific heat measurements~\cite{ramirez} on Y$_2$BaNi$_{1-x}$Zn$_x$O$_5$
which show a Schottky anomaly consistent with free $s=1$.

Another possibility for doping exists in Y$_2$BaNiO$_5$. Carriers,
in the form of holes, can be added to the system by replacing the
{\it off-chain} Y$^{3+}$ with Ca$^{2+}$.
DiTusa et. al~\cite{ditusa} show,
from polarized x-ray absorption spectra, that the added holes are
localized on the oxygen {\it between} the Ni atoms. The addition
of a hole on the oxygen sites could effectively turn these into
$s=1/2$ spins, if strongly localized. Ca doping should then have
significant effects on the spectrum.
Powder averaged inelastic neutron scattering have been performed
on Y$_{2-x}$Ca$_x$BaNiO$_5$~\cite{ditusa} showing that
Ca doping produces states in the Haldane gap. From
analysis of the spectral weight per impurity, DiTusa et
al.~\cite{ditusa} have proposed that the magnetic disturbance
caused by the Ca doping introduces either a triplet or a quartet
of subgap bound states.

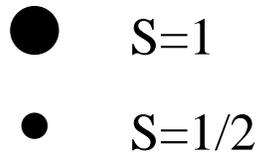
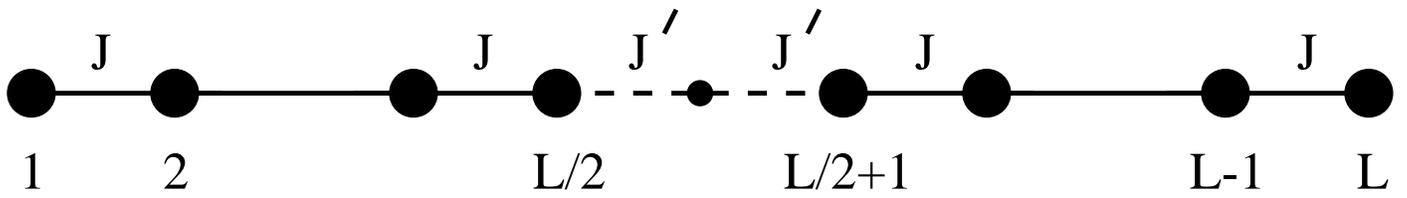
\begin{figure}
\caption{Schematic picture of the two different perturbations.
In a) is shown the perturbed bond with coupling strength $J'$
at the center of a $s=1$ chain with coupling $J$. In b) is shown
the $s=1/2$ impurity coupled with coupling strength $J'$ to
the rest of the $s=1$ chain.
}
\label{fig:chains}
\end{figure}

In order to address the experimental results on doped
$s=1$ spin chains we therefore consider the simple model Hamiltonian
\begin{equation}
H=J\sum_{i=1}^{L/2-1}\{{\bf S}_{i}\cdot{\bf S}_{i+1}\}+
J\sum_{i=L/2+1}^{L-1}\{{\bf
S}_{i}\cdot{\bf S}_{i+1}\}+H_{\rm imp}.
\label{eq:h}
\end{equation}
Here the single ion anisotropy is neglected for simplicity, and
{\bf S} represents $s=1$ spins.
The impurity Hamiltonian $H_{\rm imp}$ is taken to be either a
perturbed bond
\begin{equation}
H_{\rm imp}=J'{\bf S}_{L/2}\cdot{\bf S}_{L/2+1},
\label{eq:himpa}
\end{equation}
or a $s=1/2$ impurity, in which case we have
\begin{equation}
H_{\rm imp}=J'{\bf S}_{L/2}\cdot{\bf S'}+J'{\bf S'}\cdot{\bf S}_{L/2+1},
\label{eq:himpb}
\end{equation}
where $S'$ is the $s=1/2$ impurity spin.
Schematically the two kinds of impurities are shown in
{}Fig.~\ref{fig:chains}a,~\ref{fig:chains}b, respectively.
{}For the carrier doped compounds Eq.~(\ref{eq:h}) may not
be a good description and a
Hubbard like model may be more appropriate. However,
in themselves the Hamiltonians, Eq.~(\ref{eq:h}), deserves
attention and they should form a good starting point
also for the carrier doped compounds if the carriers are
strongly localized.

Our approach is to study the low-lying excitations in these
systems by comparing large-scale numerical results,
obtained using the density matrix renormalization group (DMRG) method,
to perturbative results, valid at strong or large coupling, based
on effective $s=1/2$ chain-end excitations generated by broken valence bonds.
In section~\ref{sec:methods} we discuss the numerical approach.
Section~\ref{sec:bonds} addresses the effects of a single
perturbed bond at the center of a chain, while we in Section~\ref{shalf}
present our results for a $s=1/2$ impurity. The bulk of our results
concerns antiferromagnetically coupled impurities, however, in
Sec.~\ref{sec:ferro.coupling}
we briefly discuss scenarios for ferromagnetically
coupled impurities.

\section{Numerical Methods}\label{sec:methods}
We use the density matrix renormalization group (DMRG) method proposed
by White~\cite{noack,white1,white2}.
{}For a detailed discussion of the DMRG method we refer
the reader to Refs.~\onlinecite{white1,white2}.
In all our runs we truncate each half-space to $m=81$ states,
and we work in a subspace
defined by the total z-component of the magnetization,
$S^z_T$, and by the parity with respect to reflection around the central
bond, or central spin, for even and odd length chains, respectively.
{}For the DMRG method we therefore denote the different states
by $S^{z\ P}_T$. Note
that, this is the z-component of the spin and {\it not} the total spin
of the state.

The idea behind the DMRG method revolves around the notion of a
subsystem, in this case half of the chain, that is strongly coupled
to the universe, here the complete chain. As usual
in quantum mechanics, the subsystem is then best described
by its density matrix, whose eigenvalues give the probability for
the subsystem to be in a given eigenstate (of the subsystem density
matrix), under the constraint that the universe (the complete chain)
is in a given pure state. Here we mean, by pure state, an eigenstate
of the Hamiltonian for the complete system. It is now clear that if we start
to effectively decouple the subsystem from the universe, by changing
the relevant coupling at the center of the chain, the standard DMRG method,
where the system is split in half at the impurity site,
will become progressively inaccurate. This can be seen in the limit
where the coupling to the universe, i. e. the central bond in the chain,
is zero, in which case the eigenstates of the density matrix and the
Hamiltonian for the
subsystem are the same and the DMRG method reduces to standard
real space renormalization which is not comparable to the strongly
coupled DMRG in precision. We therefore expect to have problems whenever
the perturbations that we consider at the center of the chain behave
in a way that effectively breaks the system in two. However, as we shall
see, this still leaves a workable region where we can obtain useful
results. We assume that these difficulties could be overcome by considering
more complicated DMRG schemes where the system is not split at the
impurity site.  However, most other schemes will reduce the
symmetry of the problem and we therefore prefer to work with the
standard method.

\section{Bond Doping}\label{sec:bonds}
In the case where the Ni based $s=1$ Heisenberg antiferromagnets are doped by
impurities that form divalent non-magnetic ions, such as Zn, Cd, and Hg,
we expect the impurities to effectively cut the Ni chains.
In the valence bond picture~\cite{aklt} this leads to free
$s=1/2$ at the neighboring sites. In this section we consider the more general
case where the central bond in an open chain is different from the rest
of the chain, although still antiferromagnetic. The Hamiltonian describing
the system is Eq.~(\ref{eq:h}) and (\ref{eq:himpa}). For the experiments
performed with non-magnetic impurities the relevant
situation is that of an extremely weak bond.
This can be seen in the experiments of Glarum et al.~\cite{glarum}
who interpret their results in terms of {\it free} $s=1/2$ chain-end spins.

\subsection{Weak Coupling}\label{sec:weak.link}
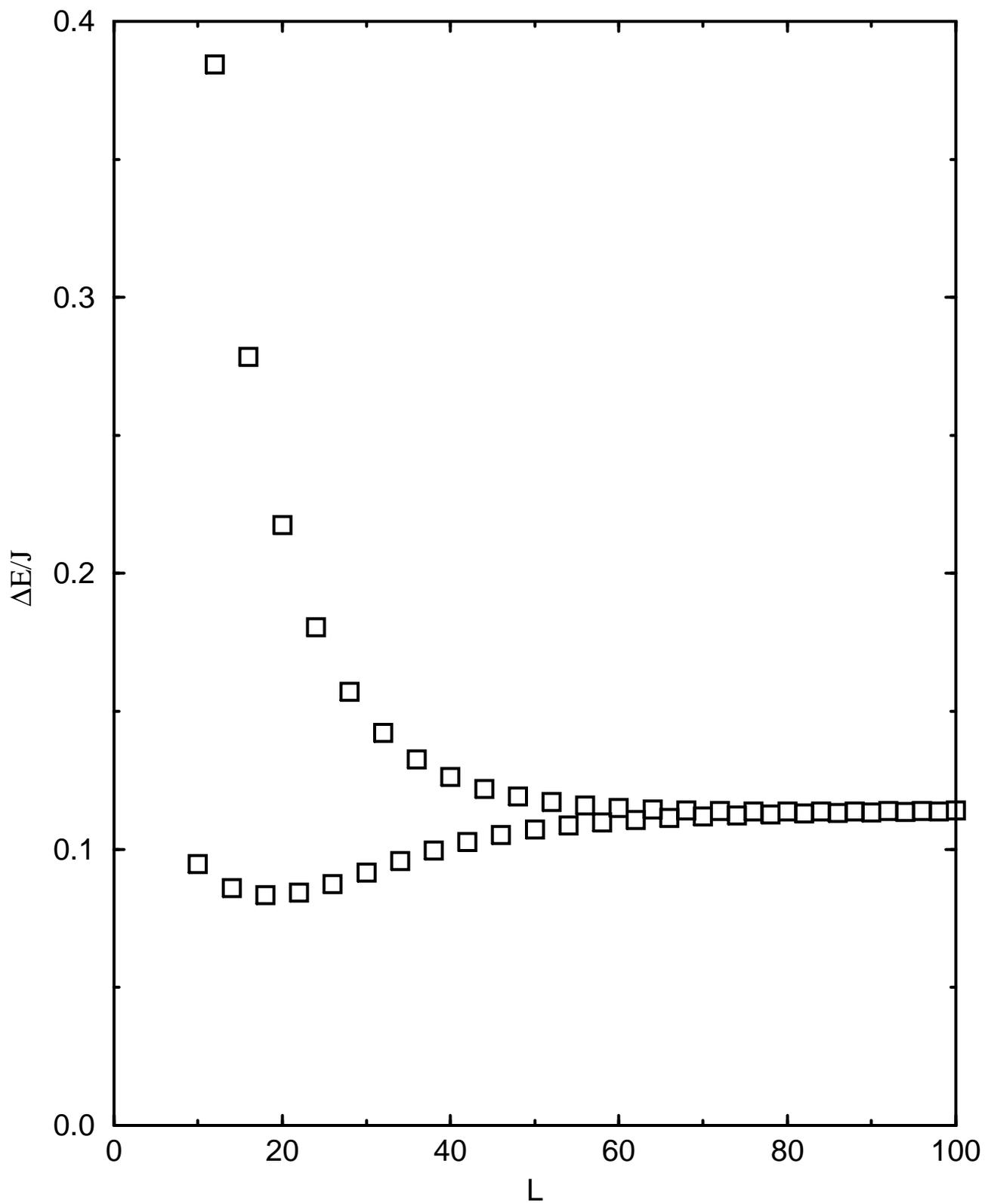
\begin{figure}
\caption{The energy gap $\Delta E(L)=E_{2^+}(L)-E_{1^-}(L)$ as
a function of chain length, $L$, for a weak bond, $J'/J=0.1$.
The limiting value of the gap is 0.11405.
}
\label{fig:gapj.1}
\end{figure}

If the bond at the center of the chain is weak
enough we can view the problem as two weakly
coupled open chains of half the length. At the end of
an open $s=1$ chain we expect to find an
effective $s=1/2$ excitation, this we shall refer to as a chain-end
spin.
Thus, for sufficiently long chains,
the low-lying excitations can be described by an
effective Hamiltonian of two weakly coupled
chain-end spins. This should be valid once the coupling of the
chain-end spins across the impurity bond is larger than the
coupling {\it along} the chain segment. The latter is known to
decrease exponential with the length of the chain
segment~\cite{hagiwara,partha,anis}, and should thus
quickly become negligible. Experimentally this is the relevant
situation for NENP, since it is difficult to obtain high Zn doping
levels~\cite{glarum}, and the average chain segment is
therefore likely to be long. Y$_2$BaNiO$_5$ offer
the possibility of higher doping levels.
If we denote by ${\cal P}^{(1/2)}$ the projection onto the
$s=1/2$ subspace, we can write
${\cal P}^{(1/2)}{\bf S}_{s=1}{\cal P}^{(1/2)}=\alpha{\bf S''}_{s=1/2}$.
We need to determine $\alpha$. Fortunately,
an estimate can easily be obtained by looking at the expectation
value one of the spins at the {\it end} of a {\it free} chain.
If we by $|1>$ denote the state that has both the chain-end spins
polarized in the ``up"-state we find
\begin{eqnarray}
<1|{\bf S}^{z}_{s=1}|1>& \simeq &
<1|{\cal P}^{(1/2)}{\bf S}^z_{s=1} {\cal P}^{(1/2)}|1>\nonumber\\
& = & \alpha <1|{\bf S}^{z\ \prime\prime}_{s=1/2}|1>\ =\ \frac{\alpha}{2}.
\end{eqnarray}
Previous work~\cite{white1,anis} have established that
$<1|{\bf S}^{z}_{s=1}|1>\simeq 0.5320$
at the end of a free chain, and thus $\alpha\simeq 1.0640$.
The effective Hamiltonian can then be written as
\begin{equation}
H_{\rm eff}= \alpha^2J'{\bf S''}_{L/2}\cdot{\bf S''}_{L/2+1}
\ \ J'\ll J,
\label{eq:heff.weak}
\end{equation}
where ${\bf S''}$ denotes a $s=1/2$ spin. We then immediately see that
the low energy spectrum will be given by a ground-state singlet and
a higher lying triplet, the splitting between the two given by
\begin{equation}
\Delta E = \alpha^2 J',\ J'\ll J.
\label{eq:deweak}
\end{equation}
This corresponds to a bound state in the Haldane gap, $\Delta$. For the
isotropic antiferromagnetic Heisenberg model that we are studying here
we have $\Delta=0.4105J$ in the absence of any impurities.
In Fig.~\ref{fig:gapj.1} we show the energy gap between the two lowest
lying states for a chain that has a weak bond of $J'=J/10$ in the middle
of an otherwise isotropic chain. Since we have two essentially free
$s=1/2$ chain-end spins at each end of the open chain, the ground-state
becomes four-fold degenerate. For finite chain lengths we expect these
two chain-end spins to combine into a ground-state singlet, $0^+$, with
an exponentially low-lying triplet, $1^-$~\cite{bose,anis}.
In the thermodynamic limit the triplet and the singlet become degenerate.
Thus, if we want to calculate the spectrum caused by the impurity
we need to look at higher excited states. The resulting
gap, which remain finite in the thermodynamic limit, can be calculated
between the $2^+$ and $1^-$ states, where we for computational convenience
take the $1^-$ state to represent the ground-state since they become
degenerate in the thermodynamic limit.
As is clearly evident in Fig.~\ref{fig:gapj.1},
the gap converges quickly to a
value of $\Delta E=0.1141$.
With $J'=0.1J$ we find that Eq.~(\ref{eq:deweak}) gives $\Delta
E=0.1132$ in excellent agreement.
The good agreement between the perturbative results
and the DMRG results lends further support to the picture of
effective $s=1/2$ chain-end spins.

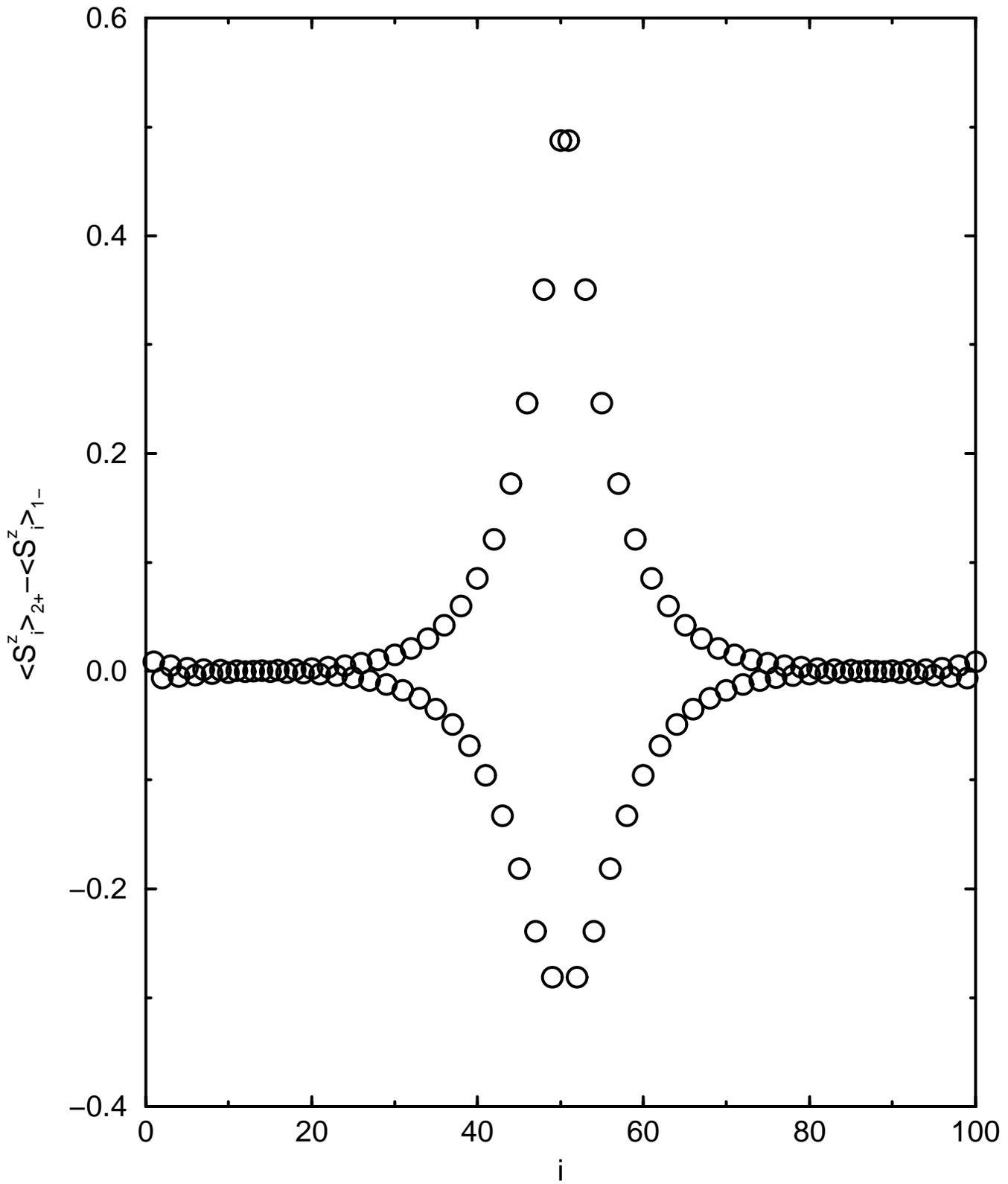
\begin{figure}
\caption{The bound state in real space for a chain with $L=100$.
$<S^z_i>_{2^+}-<S^z_i>_{1^-}$ for a weak bond, $J'/J=0.1$.
Clearly visible is
the localized state around the impurity site.
}
\label{fig:szj.1}
\end{figure}

The DMRG method is convenient for studying correlations in real-space.
In Fig.~\ref{fig:szj.1} we show the expectation value of the
z-component of the spin along a 100 site chain for $J'=0.1J$.
In the $2^+$ state we have
effectively polarized all four $s=1/2$ chain-end spins at sites
1, 50, 51, and 100 in
the ``up" direction. In order to focus attention on the effects
of the impurity we have subtracted off $<S^z_i>_{1^-}$. In the $1^-$
state the chain-end spins at site 50, 51 are combined into a singlet,
by subtracting off $<S^z_i>_{1^-}$ we therefore cancel out
the effect of the chain-end spins at site 1 and 100.
One sees that the two chain-end spins surrounding the perturbed bond
have effectively formed a localized state.

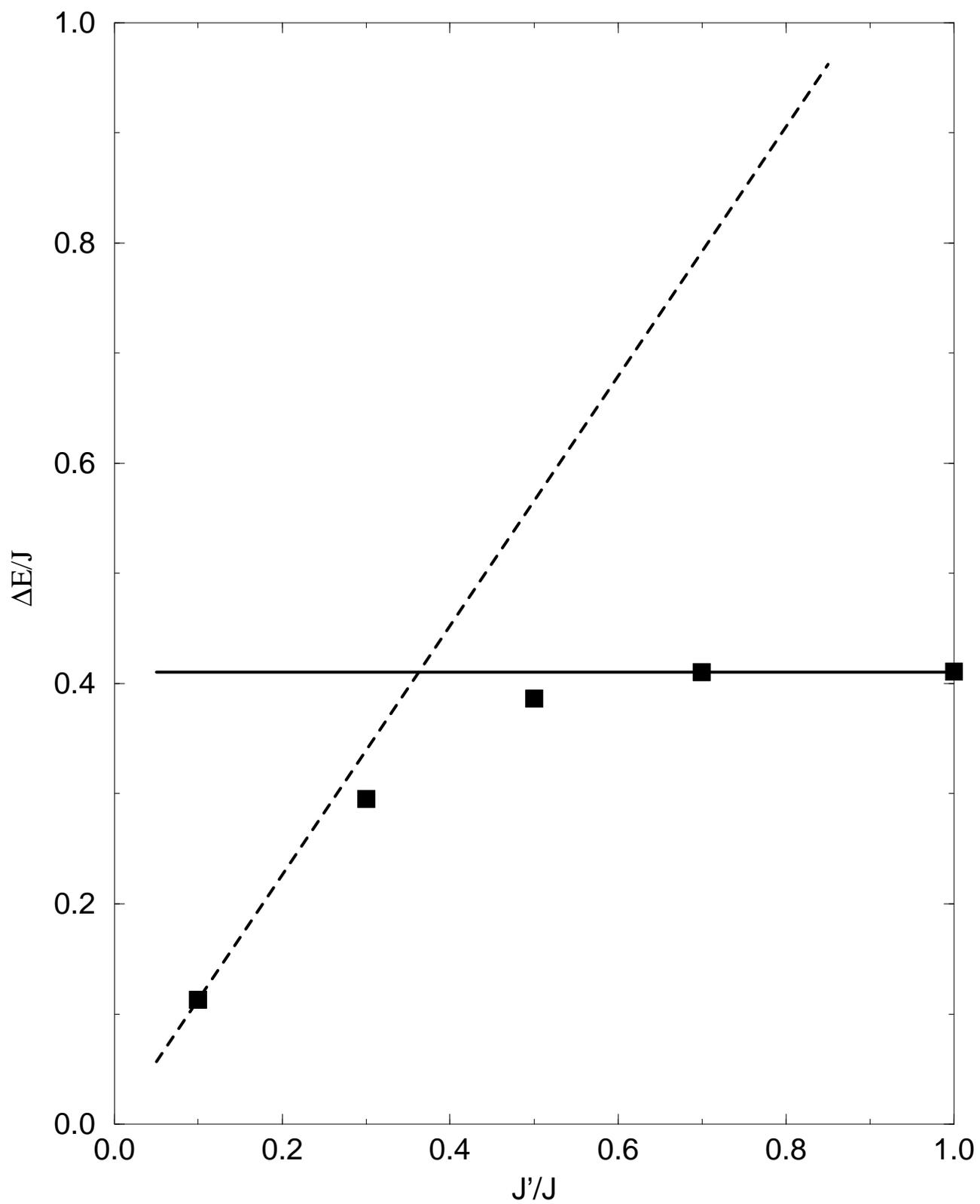
\begin{figure}
\caption{The energy gap, $\Delta E=E_{2^+}-E_{1^-}$ (solid squares), as
a function of the weak perturbed bond, $J'/J$.
The solid line is the isotropic Haldane gap, $\Delta/J=0.41050$.
The dashed line indicates the perturbative result,
Eq.~(\protect\ref{eq:deweak}).
}
\label{fig:weakgap}
\end{figure}

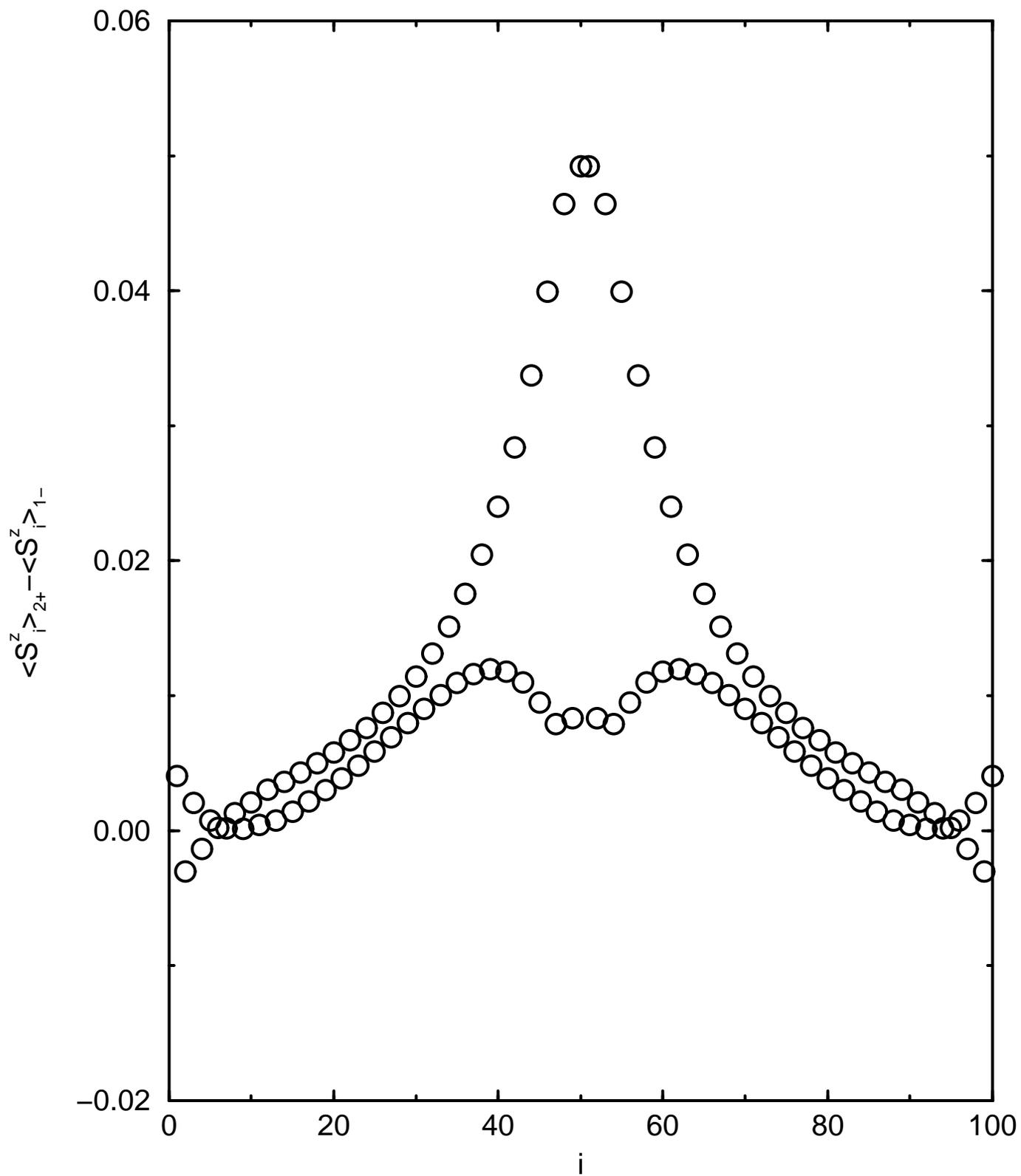
\begin{figure}
\caption{
$<S^z_i>_{2^+}-<S^z_i>_{1^-}$ as function of chain index $i$
for a weak bond, $J'/J=0.7$, close to the critical coupling.
}
\label{fig:szj.7}
\end{figure}

Using the DMRG method we can now map out the energy of this
bound state as a function of the weak coupling, $J'$. Due to
the problems mentioned in Section~\ref{sec:methods} it is not
possible to treat values of $J'$ below $J/10$ within the standard DMRG.
However, since the impurity state is localized, the convergence
to the thermodynamic limit is very fast as can be seen in
{}Fig.~\ref{fig:gapj.1}.
Our results are shown in Fig.~\ref{fig:weakgap}.
The filled squares are the numerical results, the solid line
the Haldane gap for the isotropic chain, $\Delta/J=0.41050$, and
the dashed line indicates the result from Eq.~(\ref{eq:deweak}).
{}From Fig.~\ref{fig:weakgap} it is seen that only for $J'<0.7J$
does a bound state below the isotropic Haldane gap exist. Around
$J'=0.7J$ the bound state seen in Fig.~\ref{fig:szj.1} delocalizes
and spreads out over the entire chain. This is shown in Fig.~\ref{fig:szj.7}
where we plot $<S^z_i>_{2^+}-<S^z_i>_{1^-}$ as a function of
chain index $i$ for a 100 site chain with $J'=0.7J$.
Clearly this state is not localized and it resembles what one
finds for the isotropic chain~\cite{bose}. We therefore conclude that
it is necessary to weaken the central bond below
a critical value $J'_{c\ \rm weak} \approx 0.7J$
in order to create a bound state.

\subsection{Strong Coupling}\label{sec:strong.link}
We now turn to a discussion of the opposite limit where the
central bond is very strong, $J'\gg J$. For the antiferromagnetic
coupling that we consider here, the two
neighboring spins will be contracted into a singlet. In the
limit $J'\rightarrow\infty$ the
chain will be severed as was the case for a weak bond. We now
want to derive an effective Hamiltonian valid for $J'\gg J$.
As a starting point it is convenient to consider the
case where we only have four spins described by the Hamiltonian:
\begin{equation}
H = J{\bf S}_1\cdot {\bf S}_2 + J'{\bf S}_2\cdot
{\bf S}_3 + J{\bf S}_3\cdot {\bf S}_4.
\label{eq:hstrong}
\end{equation}
We do perturbation theory in $J/J'$.  Note that when $J=0$ there is a
degneracy since the spins, ${\bf S}_1$ and ${\bf S}_4$ can be in
arbitrary states.  This is lifted in second order perturbation theory in
$J$.  The result of second order perturbation theory can be expressed in
terms of an effective Hamiltonian acting on the low-energy subspace, in
which ${\bf S}_2$ and ${\bf S}_3$ are projected onto the singlet state:
\begin{equation}
H_{\rm eff} = PH'{1\over E_0-H_0}H'P.
\end{equation}
Here
\begin{eqnarray}
H_0&\equiv &J'{\bf S}_2\cdot {\bf S}_3, \\
H' &\equiv & J{\bf S}_1\cdot {\bf S}_2 +J{\bf S}_3\cdot {\bf S}_4,
\end{eqnarray}
and $P$ projects onto the singlet ground-state of $H_0$,
with energy $E_0=-2J$.  Acting with $H'$ on the ground-state of $H_0$
produces the eigenstate of $H_0$ with spin 1 and energy, $E_1=-J$.
Thus:
\begin{eqnarray}
H_{\rm eff}&=&{J^2\over -J'}P({\bf S}_1\cdot {\bf S}_2+
{\bf S}_3\cdot {\bf S}_4)^2P \nonumber\\
&=& {\rm constant} -{2J\over J'}P({\bf
S}_1\cdot {\bf S}_2)({\bf S}_3\cdot {\bf S}_4)P\nonumber\\
&=&  {\rm constant} -{2J\over J'}<0|S_2^aS_3^b|0>S_1^aS_4^b,
\end{eqnarray}
where $|0>$
denotes the spin-0 ground-state of $H_0$.  Using:
\begin{equation}
<0|S_2^aS_3^b|0> = {1\over 3}\delta^{ab}<0|{\bf S}_2\cdot {\bf S}_3|0>
=-{2\over 3}\delta^{ab},
\end{equation}
we obtain:
\begin{equation}
H_{\rm eff}={4\over 3} {J^2\over J'} {\bf S}_1\cdot {\bf S}_4.
\end{equation}
The same calculation applies to a chain of arbitrary length.
$H_{\rm eff}$ now contains all the additional couplings which don't
involve the pair of strongly coupled spins, ${\bf S}_{L/2}$ and ${\bf
S}_{L/2+1}$ ie.:
\begin{equation}
H_{\rm eff}=J\sum_{i=1}^{L/2-2}{\bf
S}_i\cdot {\bf S}_{i+1} +J\sum_{i=L/2+2}^{L-1}{\bf S}_i\cdot {\bf
S}_{i+1}+ {4\over 3} {J^2\over J'} {\bf S}_{L/2-1}\cdot
S_{L/2+2}.
\end{equation}
This is exactly the same as the original
Hamiltonian of Eq.~(\ref{eq:h}) with two sites removed from the ends of the
chain and $J'$ replaced by $4J^2/3J'$.  Assuming this impurity coupling
is much smaller than $J$ we write an effective low-energy Hamiltonian by
proceeding along the lines of Section~\ref{sec:weak.link}.
By introducing spin-1/2 variables, ${\bf S}_{L/2-1}''$ and ${\bf
S}_{L/2+2}''$, we obtain:
\begin{equation}
H_{\rm eff}={4\over
3}{J^2\over J'}\alpha^2 {\bf S}_{L/2-1}''\cdot {\bf
S}_{L/2+2}''.
\label{eq:heff.strong}
\end{equation}
{}From this effective Hamiltonian we can then derive a simple expression
for the bound state:
\begin{equation}
\Delta E = \frac{4}{3}\alpha^2\frac{J^2}{J'},\ \ J'\gg J.
\label{eq:destrong}
\end{equation}

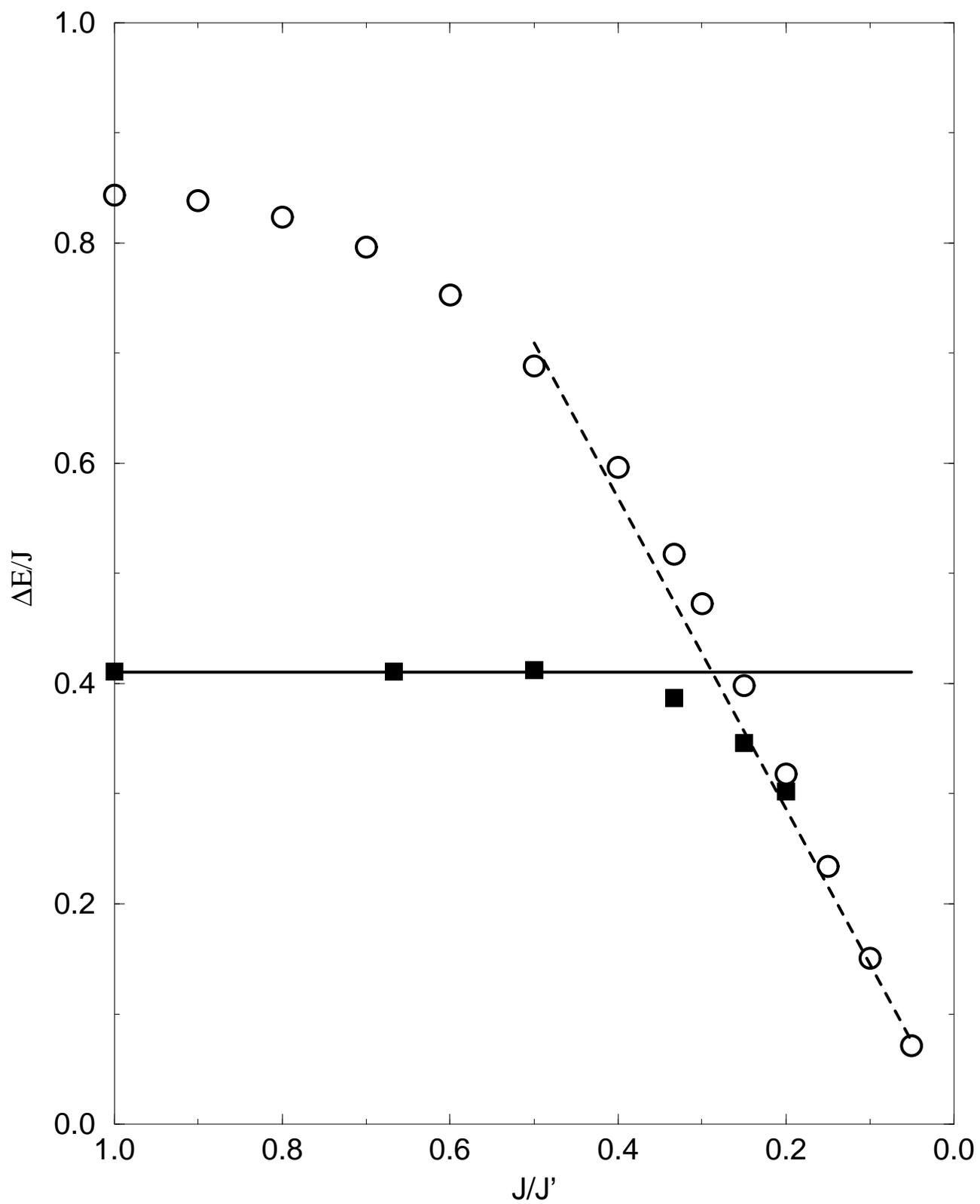
\begin{figure}
\caption{The energy gap $\Delta E=E_{2^+}-E_{1^-}$ as
a function of the strong perturbed bond, $J/J'$.
The solid squares are the DMRG results.
The solid line is the isotropic Haldane gap, $\Delta/J=0.41050$.
The dashed line indicates the perturbative result,
Eq.~(\protect\ref{eq:destrong}).
The open circles denote exact diagonalization results of
Eq.~(\protect\ref{eq:hstrong}).
}
\label{fig:destrong}
\end{figure}

We have calculated the gap $\Delta E= E_{2^+}-E_{1^-}$ as a function
of the coupling, $J/J'$, to the $s=1/2$ impurity spin. Our results,
extrapolated to the thermodynamic limit, are shown in Fig.~\ref{fig:destrong}.
The solid line indicates the Haldane gap, $\Delta/J=0.41050$, for the pure
chain. The solid squares are the numerical results
obtained using the DMRG method.
The dashed line denotes the perturbative result Eq.~(\ref{eq:destrong}).
The open circles indicate a direct diagonalization of Eq.~(\ref{eq:hstrong})
with ${\bf S}_1$ and ${\bf S}_4$ spin one-half.
As explained in Section~\ref{sec:methods} we are limited to
study $J/J'\ge 0.25$.
{}For $J/J'=0.5$ the limiting value of the gap is very close to the value
of the isotropic chain, $\Delta$, and for $J/J'\le 0.5$
there appears to be no bound state in the Haldane gap. For smaller
$J/J'$ a bound state emerges, and we see good agreement between the
numerical results and the effective Hamiltonian Eq.~(\ref{eq:heff.strong}).
As was the case for a weak bond we find a critical coupling,
$J/J'_{c\ \rm strong}$,
but now with an approximate value of
$J/J'_{c\ \rm strong}\approx 0.5$ above which no bound state occurs.

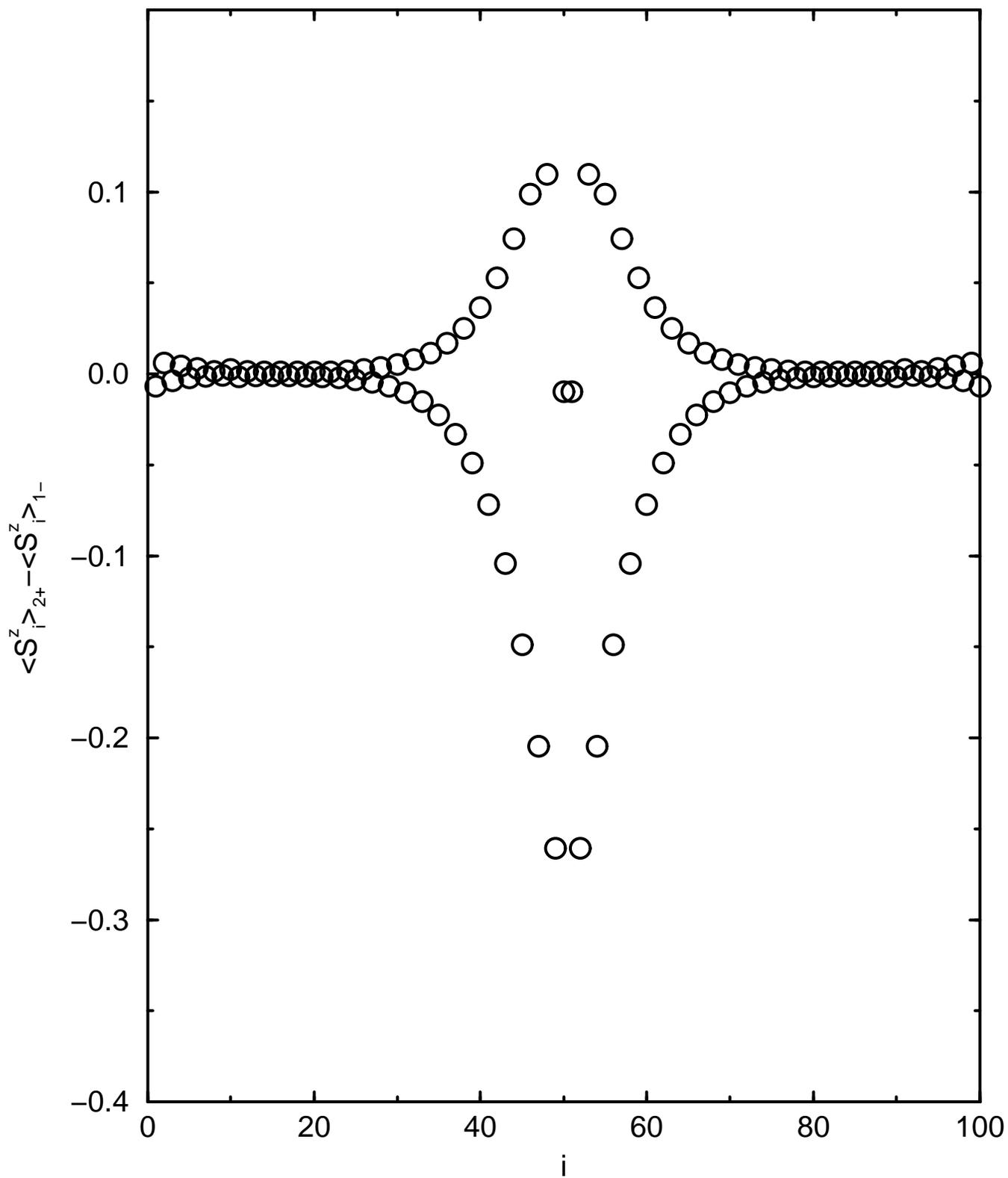
\begin{figure}
\caption{The bound state in real space for a chain with $L=100$.
$<S^z_i>_{2^+}-<S^z_i>_{1^-}$ as a function $i$
for a strong bond, $J/J'=0.25$. The quenched singlet right at the
center of the chain is clearly visible.
}
\label{fig:szj4.0}
\end{figure}

The picture of a quenched singlet in the middle of the chain
weakly coupled to two chain-end spins on either side, can be substantiated
by calculating the expectation value of the z-component of the spin
along the chain. This is shown in Fig.~\ref{fig:szj4.0} for $L=100$,
$J/J'=0.25$. Again
we have subtracted off the $1^{-}$ state configuration. The quenched
singlet is clearly visible at the center of the chain and we see
a large signal from the two polarized chain-end spins at $i=L/2-1$
and $i=L/2+2$. Since one site has effectively been removed from each
of the two half-chains Fig.~\ref{fig:szj4.0} appears inverted
with respect to Fig.~\ref{fig:szj.1}. At the point where the
bound state disappear we again find that the expectation
value $<S^z_i>_{2^+}-<S^z_i>_{1^-}$ spreads out over the entire chain.

We may think of a magnon as behaving, in many
respects, like an ordinary particle of mass $m=\Delta /v^2$, where $v$
is the spin-wave velocity. Thus it may, at first, seem surprising that
no bound state exists for a weak perturbation of the coupling,
$J'\approx J$ since there is a well-known theorem of one-dimensional
quantum mechanics which states that an attractive potential always
produces a bound state, no matter how weak it is~\cite{simon}.  Our
result is not surprising if one looks in a bit more detail at the
mapping of the spin problem into an effective one-particle quantum
mechanics problem.

We represent the spin operators at long length-scales in terms of the
non-linear $\sigma$ (NL$\sigma$) model field, $\bbox{\phi}$ (obeying the
constraint $|\bbox{\phi}|^2=1$) and the associated canonical momentum,
${\bf  l}=(1/vg)\bbox{\phi}\times \partial \bbox{\phi}/\partial t$, as:
\begin{equation}
{\bf S}_i \approx s(-1)^i \bbox{\phi}+{\bf
l}.
\end{equation}
The Hamiltonian density becomes:
\begin{equation}
{\cal H}= {vg\over 2}{\bf l}^2+{v\over 2g}\left( {d\bbox{
\phi}\over dx}\right)^2.
\end{equation}
Acting on the vacuum, $\bbox{\phi}$
creates the triplet of single-magnon states.  A local
perturbation in the exchange couplings leads to additional local
terms in the field theory Hamiltonian density of the form:
\begin{equation}
\delta {\cal H} = vV(x)\left({d \bbox{\phi}\over dx}\right)^2
+ V'(x){\bf l}^2,
\end{equation}
where
$V(x)$ and $V(x')$ are some short-range functions.  Note that we {\it
do not} get simply $V(x)\bbox{\phi}^2$, since this term is trivial due
to the constraint, $\bbox{\phi}^2=1$.

We sometimes
approximate the NL$\sigma$ model by a free massive field theory model
with Hamiltonian density~\cite{FB}:
\begin{equation}
{\cal H} \approx {v\over
2}{\bf \Pi}^2+ {v\over 2}\left( {d\bbox{\phi}\over dx}\right)^2
+{\Delta \over 2v}\bbox{\phi }^2,
\end{equation}
where the constraint on $\bbox{\phi}$ is now relaxed and ${\bf \Pi}$
is the associated cononical momentum.  This gives the dispersion
relation:
\begin{equation}
E=\sqrt{\Delta^2+(vk)^2}.
\end{equation}
(We have
shifted the wave-vector by $\pi$.)   We may easily treat the
additional impurity term, $vV(x)(d\bbox{\phi}/dx)^2$.  This gives
the Schr\"odinger equation:
\begin{equation}
-(1/2)\psi '' -[V(x)\psi ']' = mE\psi .
\end{equation}
We obtain this equation for each of the three magnon polarizations.
Note that this {\it does not} correspond to a standard potential
scattering term.  We could obtain a potential scattering term from a
term in the NL$\sigma$ model Hamiltonian of the form $V(x)(\phi^z)^2$ or
$V(x)l^z$.  However, these terms break the $SU(2)$ symmetry and
therefore can't appear in the Heisenberg model.  The latter term
corresponds to a local magnetic field.  The former could arise if we
have a crystal field term in addition to the Heisenberg exchange.

The peculiar derivative potential term does not necessarily lead to a
bound state if it is weak.  To demonstrate this we consider the case
of a step function:
\begin{eqnarray}
V(x) &=&0 \ \ \ (|x|>a)\nonumber \\
V(x) &=& -V_0 \ \ \ (|x|<a).
\end{eqnarray}
Solutions to the Schr\"odinger equation will have discontinuous $\psi '$
at $x=\pm a$. Defining $\psi '(a^\pm )$ to be the limit of $\psi '(x)$
as $x\to a$ from above or below respectively and similarly for $\psi
'(-a^\pm )$ we obtain:
\begin{eqnarray}
\psi '(a^+) &=& \psi '(a^-)[1-2V_0]\nonumber \\
\psi '(-a^-) &=& \psi '(-a^+)[1-2V_0].\end{eqnarray}
The important thing to notice is that if $V_0<1/2$, the derivative
does not change sign at the discontinuity.  There can then be no
solution which vanishes at $x\to \pm \infty$.  This follows since,
assuming an even wave-function, it must have the form  $\cosh
[\sqrt{2|E|m/(1-2V_0)}x$  for $|x|<a$ which is increasing as we
move away from $0$.  It has to turn into $\exp [-\sqrt{2|E|m}|x|]$ for
$|x|>a$.  This can't happen without a change in sign of the
derivative.  [$\psi$ itself must be continuous.]  Alternatively, we
could assume an odd wave-function but we have the same problem.  If
$V_0>1/2$, there is a solution.  It is an odd function of $x$; a sine
function for $|x|<a$. The binding energy is determined by:
\begin{equation}
\tan \sqrt{2|E|m\over 2V_0-1}a
=\sqrt{2V_0-1}.
\end{equation}
This has a solution for all $V_0>1/2$.
As $V_0\to 1/2^+$, we get:
\begin{equation} |E|m\to
(2V_0-1)^2/2a^2.
\end{equation}

This provides a counter-example to the statement that there is always
a bound state for a weak derivative potential term.  On the other hand,
there is a rigorous proof of the fact (see, for example Ref.
\onlinecite{simon}) for an ordinary potential term. This argument is
admittedly quite incomplete since we haven't analyzed the other,
$V(x){\bf l}^2$ term, nor dealt carefully with the passage from the
NL$\sigma$ model to the free model.  (This additional term seems to
give a type of non-linear Schr\"odinger equation.)  Ignoring these
complications, and assuming that the step function result is
universal, the above argument suggests that as $J'\to J'_c$ the
critical coupling where the bound state disappears, the binding energy
behaves as:
\begin{equation}
E_b \propto (J'-J'_c)^2.
\end{equation}
This is not inconsistent with our numerical results, but substantially
more numerical work for $J'$ close to $J'_c$ would be needed in order
to verify this dependence.

Our findings in this section are in disagreement with a
recent Schwinger-Boson calculation~\cite{lu}.

\section{$s=1/2$ Impurities}\label{shalf}
We now turn to a discussion of $s=1/2$ impurities in a $s=1$
AF Heisenberg chain. Experimentally two scenarios have been studied.
{}First, Cu$^{2+}$ ($s=1/2$) was substituted for Ni$^{2+}$ ($s=1$)
in NENP~\cite{hagiwara}.
Here, the introduced $s=1/2$ impurity structurally takes the same place
as the original Ni ion.  Experimentally
one finds a weak {\it ferromagnetic} coupling
to the neighboring chain-end spins~\cite{hagiwara}.
Secondly, carriers were added to Y$_2$BaNiO$_5$ by substituting
Ca for Y. DiTusa et al.~\cite{ditusa} argue that the holes
become localized on the O sites between the Ni ions.
Here the situation is different since the introduced $s=1/2$ impurity
does {\it not} take the place of a $s=1$, instead, it is placed
{\it between} two Ni ions. Thus, the superexchange between the two
Ni ions is broken and presumably replaced by a {\it direct} exchange
to the $s=1/2$, $J'$, which is likely to be large compared to $J$.
Aharony et al.~\cite{aharony} have argued that for
La$_{2-x}$(Sr,Ba)$_x$CuO$_{4-y}$
the analogous coupling should be large, $|J'|\gg|J|$. However,
the appropriate sign and value of $J'$ for
a simple linear chain Heisenberg model capable of describing
the experiments on Y$_{2-x}$Ca$_x$BaNiO$_5$~\cite{ditusa} is
an open question. In the following we shall limit our
discussion to the case of an antiferromagnetic coupling, $J'>0$.

\subsection{Weak Coupling}\label{sec:shalf.weak}
{}Following the approach in Section~\ref{sec:weak.link}, for
a weak bond at the center of the chain,
we can try to find an effective Hamiltonian for
a weakly coupled $s=1/2$ impurity at the center of the chain.
In this case we again view the sites $L/2$ and $L/2+1$ of the $s=1$
chain as being occupied by chain-end spins that effectively behave as if
they were spin one-half. By projecting onto the $s=1/2$ subspace
we represent the chain-end spins
at these two sites by $\alpha {\bf S''}$, where ${\bf S''}$
is now a real spin one-half.
We then obtain an effective low energy Hamiltonian. With ${\bf S'}$
representing
the $s=1/2$ impurity spin we have:
\begin{equation}
H_{\rm eff}=\alpha J'[{\bf S''}_{L/2}\cdot {\bf S'}+
{\bf S'}\cdot {\bf S''}_{L/2+1}].
\label{eq:heff3}
\end{equation}
The spectrum for this effective Hamiltonian consists
of a ground-state doublet and an excited doublet and quadruplet.
The ground-state doublet has $S^P=1/2^+$, and energy $E=-\alpha J'$.
The excited doublet, $S^P=1/2^-$, has energy $E=0$.
{}Finally, the quadruplet, $S^P=3/2^+$, has energy $\alpha J'/2$.
{}From this effective Hamiltonian we see
that we should have {\it two} low-lying
levels with energies in the Haldane gap
with energies given by
\begin{equation}
\Delta E_{1/2^-} = \alpha J',
\label{eq:de.shalf.weak.doublet}
\end{equation}
and
\begin{equation}
\Delta E_{3/2^+}=\frac{3}{2}\alpha J'.
\label{eq:de.shalf.weak.quadruplet}
\end{equation}
{}For a weak {\it ferromagnetic} coupling we would have obtained
$\Delta E_{1/2^-} = \alpha J'/2,\ \ \ \Delta E_{1/2^+}=3\alpha J'/2$.

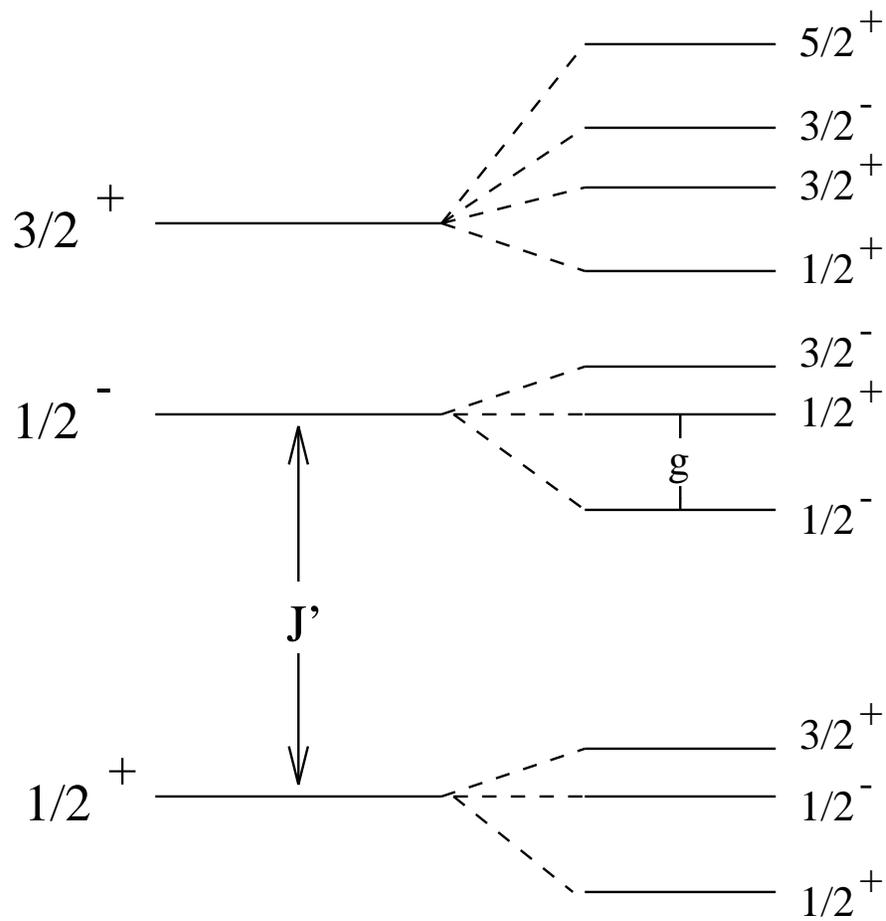
\begin{figure}
\caption{The low-lying levels for a weakly antiferromagnetically
coupled $s=1/2$ in a spin-one chain.
}
\label{fig:levels}
\end{figure}

In order to observe these two levels with the DMRG method we need to determine
the relevant quantum numbers for the complete system. We begin by considering
the case where the two half-chains surrounding the $s=1/2$ impurity are
of {\it even} length. We consider the situation where the
chain-end spins at sites $L/2$ and $L/2+1$ are coupled weakly with a coupling
$g$, $g\ll J'$,
to the chain-end spins at the sites $1$ and $L$, respectively. Since the
chain segments have an {\it even length} this coupling, $g$,
is {\it antiferromagnetic}~\cite{anis}. It is now straight forward to obtain
the level diagram shown in Fig.~\ref{fig:levels}a. If the two
half-chains have an {\it odd length} the coupling, $g$,
is {\it ferromagnetic}~\cite{anis}
and we obtain the level diagram shown in Fig.~\ref{fig:levels}b. Note
that, since we in either case have an even number of valence bonds
we do not get a change of parity from the rest of the chain.
Thus, we see that in the thermodynamic limit the ground-state
is eight-fold degenerate with an eight-fold
and sixteen-fold degenerate level above it. In our implementation
of the DMRG method we cannot use the total spin as a good quantum number
but only the total z-component. In order to determine the two levels
given by Eqs.~(\ref{eq:de.shalf.weak.doublet})
and (\ref{eq:de.shalf.weak.quadruplet})
we then see from Fig.~\ref{fig:levels}
that we need to calculate the energy differences
$E_{3/2^-}-E_{3/2^+}$ and $E_{5/2^+}-E_{3/2^+}$. Only for $L/2$ odd
is $3/2^+$ the true ground-state. The Haldane gap to bulk excitations
(i. e. excitations not localized at the center and/or ends of the
chain)
will be between the $5/2^-$ and $3/2^+$ states.

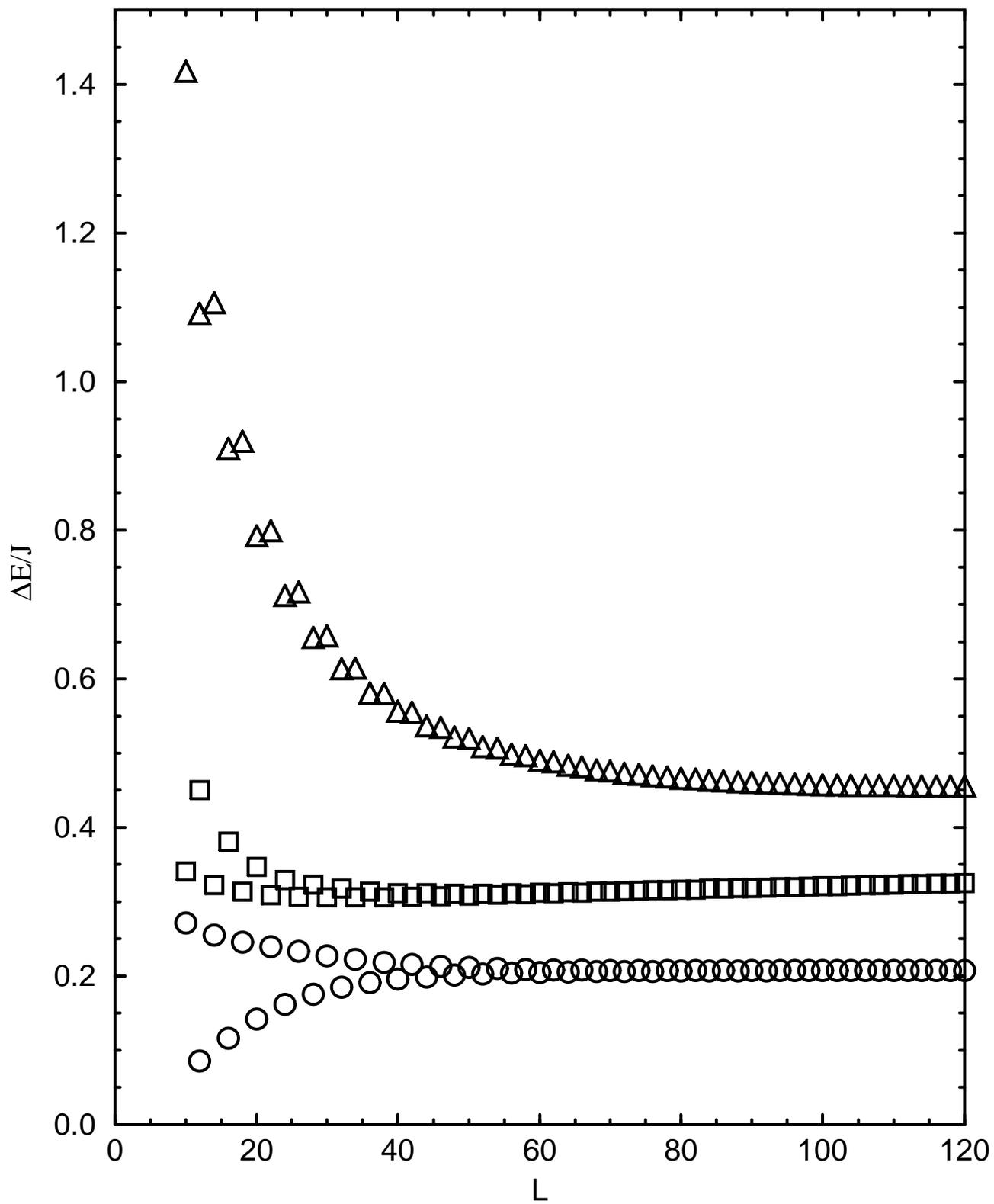
\begin{figure}
\caption{The energy gap, $\Delta E(L)$, as a function
of chain length $L$, for a $s=1$ chain with a $s=1/2$ impurity.
The coupling to the $s=1/2$ is $J'=0.2J$. The symbols
denote the gaps: $E_{5/2^-}(L)-E_{3/2^+}(L)$ (triangles) to the state
$5/2^-$ (Haldane gap), $E_{5/2^+}(L)-E_{3/2^+}(L)$ (squares) to the state
$5/2^+$, and $E_{3/2^-}(L)-E_{3/2^+}(L)$ (circles) to the state $3/2^-$.
}
\label{fig:gap.shalf.j0.2}
\end{figure}

In Fig.~\ref{fig:gap.shalf.j0.2} we show
the energy gap, $\Delta E(L)$, as a function
of chain length $L$, for a $s=1$ chain with a $s=1/2$ impurity.
The coupling to the $s=1/2$ impurity is $J'=0.2J$. The symbols
denote the gaps: $E_{5/2^-}(L)-E_{3/2^+}(L)$ (triangles) to the state
$5/2^-$ (Haldane gap), $E_{5/2^+}(L)-E_{3/2^+}(L)$ (squares) to the state
$5/2^+$, and $E_{3/2^-}(L)-E_{3/2^+}(L)$ (circles) to the state $3/2^-$.
Clearly, we have two low-lying excitations well separated from
the Haldane gap. For the $3/2^-$ state we find the gap to be
$0.2073J$, compared to $0.2128J$ from Eq.~(\ref{eq:de.shalf.weak.doublet}).
Although at large $L$ we have minor problems with the numerical
results for the $5/2^+$ state we can also in this case
estimate the gap to $\sim0.31$ which compares favorably to the
result $0.3192$ from Eq.~(\ref{eq:de.shalf.weak.quadruplet}).

It is also possible to study the spatial extension of the bound states.
However, as opposed to Section~\ref{sec:bonds}, $<S^z_i>$ for
the state $3/2^+$ is non-zero across the impurity site. It is
therefore difficult to give a quantitative interpretation of
the results since the subtraction of $<S^z_i>_{3/2^+}$ will produce more
complicated effects than just the cancellation of the contribution
from the free chain-end spins at the sites 1 and L. However,
one can still differ between extended and localized states.
We found that the two states $3/2^-$ and $5/2^+$ clearly were
localized around the impurity site for $J'=0.2J$ whereas the state
$5/2^-$ was extended.

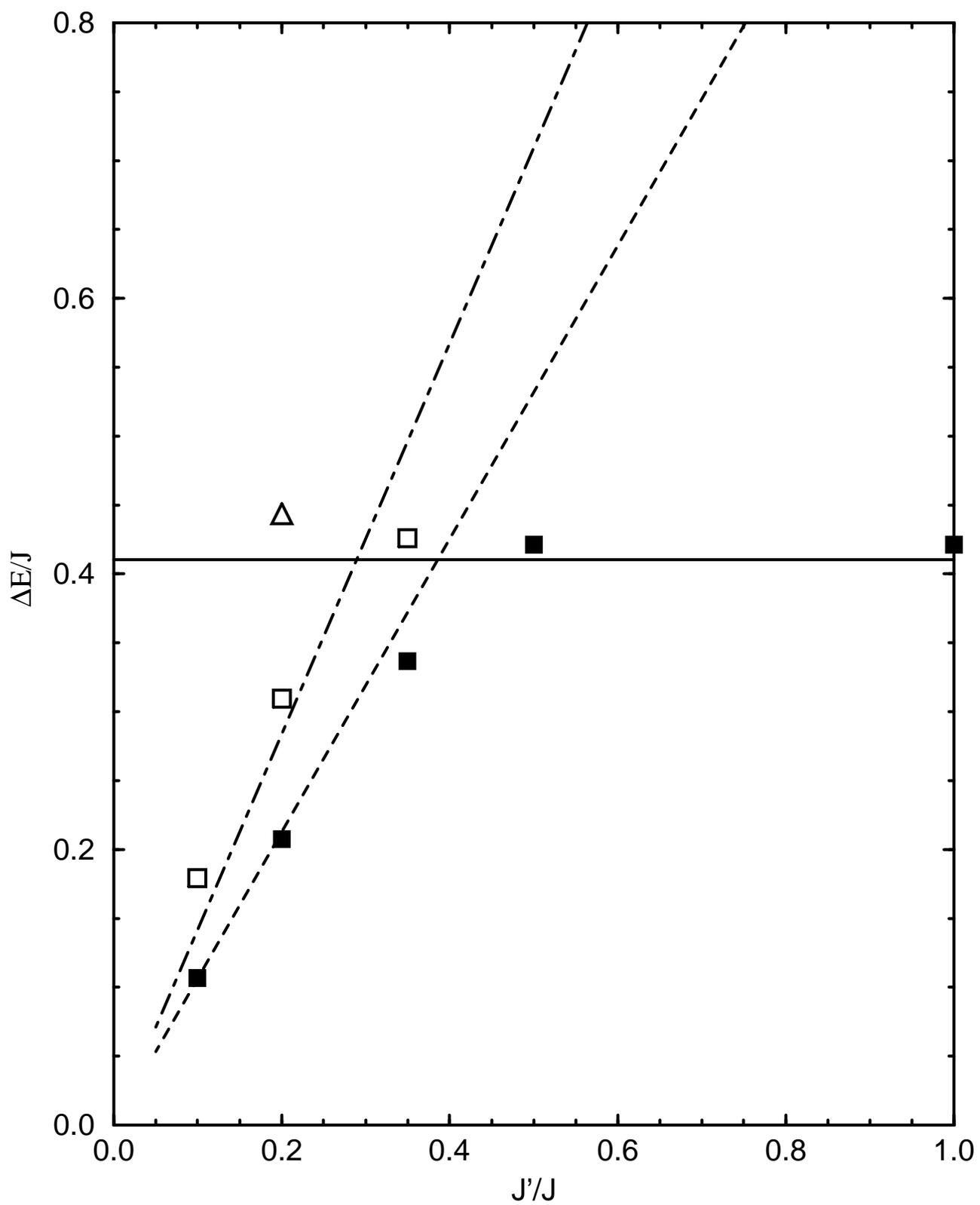
\begin{figure}
\caption{The low-lying excitations as a
function of coupling, $J/J'$, to a weakly coupled  $s=1/2$ impurity.
The symbols
denote the gaps: $E_{5/2^-}-E_{3/2^+}$ (triangles) to the state
$5/2^-$ (Haldane gap), $E_{5/2^+}-E_{3/2^+}$ (squares) to the state
$5/2^+$, and $E_{3/2^-}-E_{3/2^+}$ (filled squares) to the state $3/2^-$.
The filled squares denote the DMRG results for a 120 site chain.
The solid line is the Haldane gap, $\Delta E=0.41050J$.
The dashed line is the perturbative result ,
Eq.~(\protect\ref{eq:de.shalf.weak.doublet}),
and the dot-dashed line Eq.~(\protect\ref{eq:de.shalf.weak.quadruplet}).
}
\label{fig:shalf.weak}
\end{figure}

We can now calculate the low-lying excitations as a function
of coupling to the $s=1/2$ impurity. Our results are shown
in Fig.~\ref{fig:shalf.weak}.
The symbols
denote the gaps: $E_{5/2^-}-E_{3/2^+}$ (triangles) to the state
$5/2^-$ (Haldane gap), $E_{5/2^+}-E_{3/2^+}$ (squares) to the state
$5/2^+$, and $E_{3/2^-}-E_{3/2^+}$ (filled squares) to the state $3/2^-$.
The solid line is the Haldane gap, $\Delta E=0.41050J$.
The dashed line is the perturbative result
Eq.~(\ref{eq:de.shalf.weak.doublet}),
and the dot-dashed line Eq.~(\ref{eq:de.shalf.weak.quadruplet}).
{}For small $J'$ we see excellent
agreement with the perturbative results.
{}For $J'<J'_{c\ 1}\approx 0.45J$ a
bound state, calculated from the $3/2^-$ level,
appears, and for $J'<J'_{c\ 2}\approx 0.35 J$ also
bound state, calculated from the $5/2^+$ level, is stable.

\subsection{Strong Coupling}\label{sec:shalf.strong}

As before, we try to arrive at an effective Hamiltonian capable
of describing the low energy excitations.
Again we shall assume that the coupling to the
$s=1/2$ impurity spin, $J'$, is always positive corresponding
to an antiferromagnetic coupling.
As a starting point we consider
the three central spins ${\bf S}_{L/2},{\bf S}_{L/2+1}$
and ${\bf S'}$, the $s=1/2$ impurity.
Disregarding the rest of the chain,
these three spins will be described by a Hamiltonian
\begin{equation}
H_{\rm imp}=J'[{\bf S}_{L/2}\cdot{\bf S'}+{\bf S'}\cdot{\bf S}_{L/2+1}]
\label{eq:h0.shalf}
\end{equation}
The spectrum of $H_{\rm imp}$ is as follows:
The ground-state
is a quadruplet $S^P=3/2^+$ with $E=-3/2J'$, then a doublet
$S^P=1/2^-$ with $E=-1J'$,
another doublet $S^P=1/2^+$ with $E=0$, a quadruplet $S^P=3/2^-$ with $E=J'/2$
and finally a sextuplet $S^P=5/2^+$ with energy $E=J'$.
Since the ground-state of $H_{\rm imp}$ is a multiplet with $s=3/2$,
the effective Hamiltonian becomes that of an
effective $s=3/2$ impurity in the otherwise uniform AF $s=1$ chain.
If we by ${\cal P}^{(3/2)}$ denote the projection
onto the $s=3/2$ ground-state subspace of $H_{\rm imp}$,
we can write ${\cal P}^{(3/2)}{\bf S}_{L/2}{\cal P}^{(3/2)}
=\beta{\bf S}^{\prime\prime}_{s=3/2}$.
The effective Hamiltonian can then be written:
\begin{equation}
H_{\rm eff}=J\sum_{i=1}^{L/2-2}{\bf
S}_i\cdot {\bf S}_{i+1} +J\sum_{i=L/2+2}^{L-1}{\bf S}_i\cdot {\bf
S}_{i+1}+H_{\rm imp}(s=3/2),
\label{eq:heff4}
\end{equation}
where the term describing the $s=3/2$ impurity is given by:
\begin{eqnarray}
H_{\rm imp}(s=3/2)
&=&J[{\bf S}_{L/2-1}\cdot{\cal P}^{(3/2)}{\bf S}_{L/2}{\cal P}^{(3/2)}+
{\cal P}^{(3/2)}{\bf S}_{L/2+1}{\cal P}^{(3/2)}
\cdot{\bf S}_{L/2+2}]\nonumber\\
&=&\beta J[{\bf S}_{L/2-1}\cdot {\bf S}^{\prime\prime}_{s=3/2}+
{\bf S}^{\prime\prime}_{s=3/2}\cdot{\bf S}_{L/2+2}].
\label{eq:heff3.2}
\end{eqnarray}
Here ${\bf S}^{\prime\prime}_{s=3/2}$ denotes the effective $s=3/2$ spin.
We need to determine $\beta$.
We begin by writing down the matrices for $S^z_{L/2}$ and $S^z_{L/2+1}$
projected onto the $3/2^+$ ground-state subspace of $H_{\rm imp}$.
These have to be proportional
to a spin-3/2 representation. We find:
\begin{equation}
<\frac{3}{2}|S^z_{L/2}|\frac{3}{2}>=
<\frac{3}{2}|S^z_{L/2+1}|\frac{3}{2}>=
\left(\matrix{ {9\over {10}} & 0 & 0 & 0 \cr 0 &
{3\over {10}} & 0 & 0 \cr 0 & 0 &
 -{3\over {10}} & 0 \cr 0 & 0 & 0 & -{9\over {10}} \cr  }\right).
\end{equation}
Clearly, this is just $(3/5)S^z_{s=3/2}$, as expected. Thus,
if we project ${\bf S}_{L/2}$ and ${\bf S}_{L/2+1}$
onto the $3/2^+$ subspace we get in both cases a result
proportional to a spin-3/2 by a factor 3/5,
i.e. $(3/5){\bf S}_{s=3/2}$. Thus,
\begin{equation}
\beta = 3/5.
\end{equation}
The coupling to the $s=3/2$ impurity is therefore fixed at $J3/5$.

Let us consider the more general case where we instead
of Eq.~(\ref{eq:heff3.2})
have
\begin{equation}
H_{\rm imp}(s=3/2)
=K[{\bf S}_{L/2-1}\cdot {\bf S}^{\prime\prime}_{s=3/2}+
{\bf S}^{\prime\prime}_{s=3/2}\cdot{\bf S}_{L/2+2}],
\label{eq:heff3.2.gen}
\end{equation}
i. e. we regard the coupling to the $s=3/2$ impurity as a free parameter,
$K$. We would expect bound states to appear for $K<K_c^{(3/2)}$.
{}For very small $K$,
$K\ll K_c^{(3/2)}$, we could again assume that effective
$s=1/2$ chain-end
spins on the sites $L/2-1$ and $L/2+2$ would be a valid description,
and we would arrive at an effective low-energy Hamiltonian
consisting only of the two chain-end spins, and the $s=3/2$ impurity:
\begin{equation}
H_{\rm eff}
=\alpha K[{\bf S}^{\prime\prime}_{L/2-1}\cdot {\bf S}^{\prime\prime}_{s=3/2}+
{\bf S}^{\prime\prime}_{s=3/2}
\cdot{\bf S}^{\prime\prime}_{L/2+2}],\ \ K\ll K_c^{(3/2)}.
\label{eq:kurt}
\end{equation}
Here, $\alpha\simeq 1.064$, and ${\bf S}^{\prime\prime}_{L/2-1}, {\bf
S}^{\prime\prime}_{L/2+2}$ are $s=1/2$ spins. The low-energy Hamiltonian,
Eq.~(\ref{eq:kurt}), has the following spectrum. A ground-state doublet
with energy $E=-\alpha K 5/2$, an excited quadruplet at $E=-\alpha K$,
another quadruplet at $E=0$, and finally a sextuplet at $E=\alpha K3/2$.
{}For sufficiently small $K$ all three bound states should be stable giving
rise to energy levels in the Haldane gap given by the corresponding energy
differences, $\alpha K3/2$, $\alpha K 5/2$, and $\alpha K 4$.

Returning to the full effective Hamiltonian, Eq.~(\ref{eq:heff4}),
the question of whether bound states appear comes down
to the question of whether $J3/5$ is below or above the threshold,
$K_c^{(3/2)}$.
As we shall see below, the numerical results indicate that $J3/5$
is too large for any bound states to be stable. It is then {\it not}
possible to write down an effective low-energy Hamiltonian
based on the $s=1/2$ chain-end spins, as we did in Eq.~(\ref{eq:kurt}),
and it is necessary to work with the full effective Hamiltonian,
Eq.~(\ref{eq:heff4}), which, due to the size of the corresponding
Hilbert space, is not very useful.

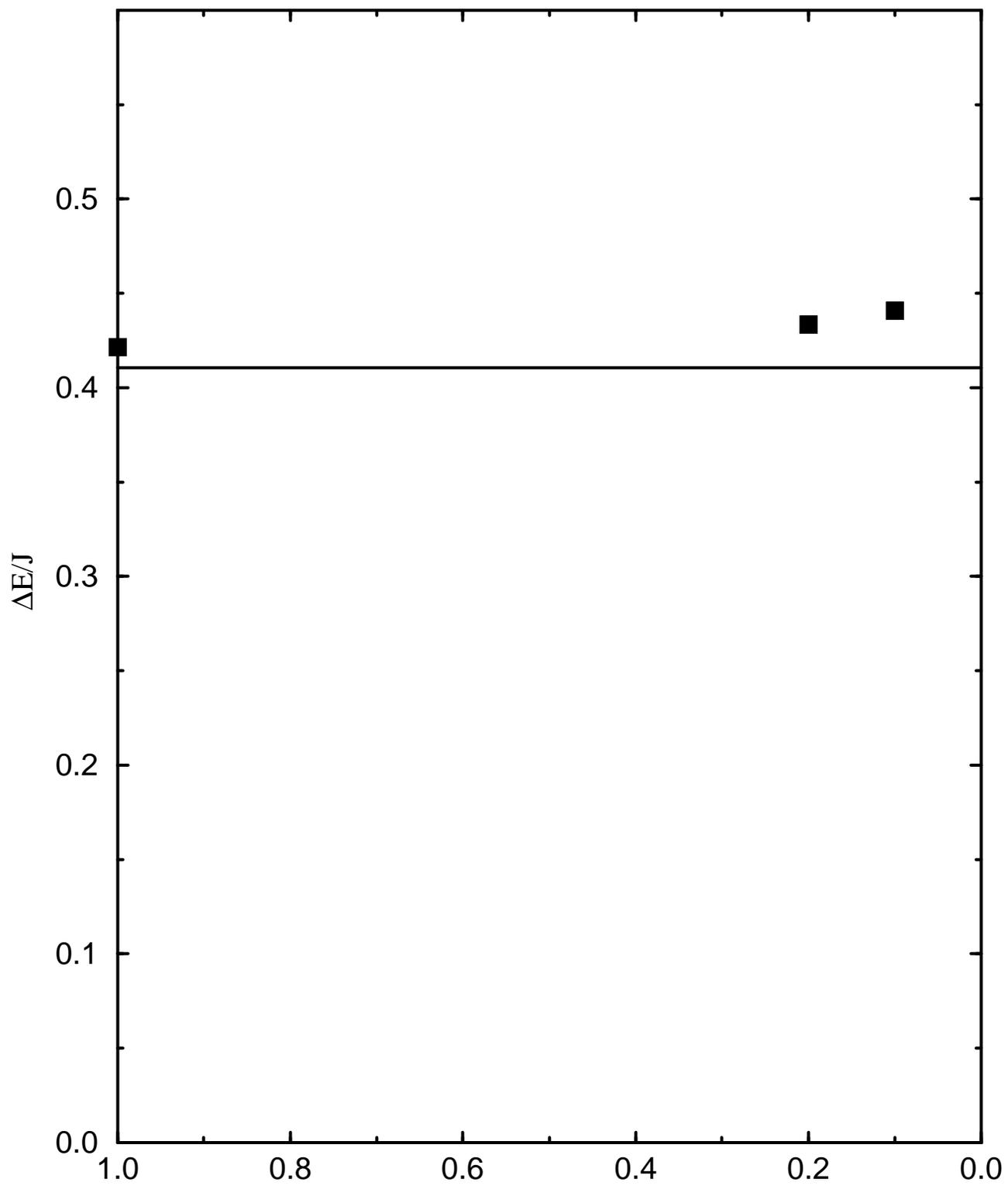
\begin{figure}
\caption{The gap, $E_{5/2^+}-E_{3/2^+}$, as a
function of coupling, $J/J'$, to the $s=1/2$ impurity.
The filled squares denote the DMRG results for a
120 site chain. No extrapolation to the thermodynamic limit
has been performed. The solid line is the Haldane gap, $\Delta E=0.41050J$.
}
\label{fig:shalf.strong}
\end{figure}

We have calculated the gap $E_{5/2^+}-E_{3/2^+}$ for $J/J'=1$, 0.2,
and 0.1 using the DMRG method. For large $J'$ this is the lowest-lying
excitation. The results are shown in
{}Fig.~\ref{fig:shalf.strong}. In this case the convergence
to the thermodynamic limit is rather slow and we have difficulty
extrapolating the results to a high accuracy, we therefore show the
unextrapolated results for a 120
site chain. Here the solid squares are the DMRG
results, and the solid line denotes the isotropic Haldane gap
of $\Delta=0.41050J$. In this case the DMRG method can
be used for even very large $J'$. As is clearly evident from
{}Fig.~\ref{fig:shalf.strong} we never observe any states
in the Haldane gap. A calculation of $E_{3/2^-}-E_{3/2^+}$
also didn't show any bound states in the Haldane gap.
We therefore conclude that the coupling to the effective
$s=3/2$ impurity, $J3/5$, is too large for any bound states
to appear.

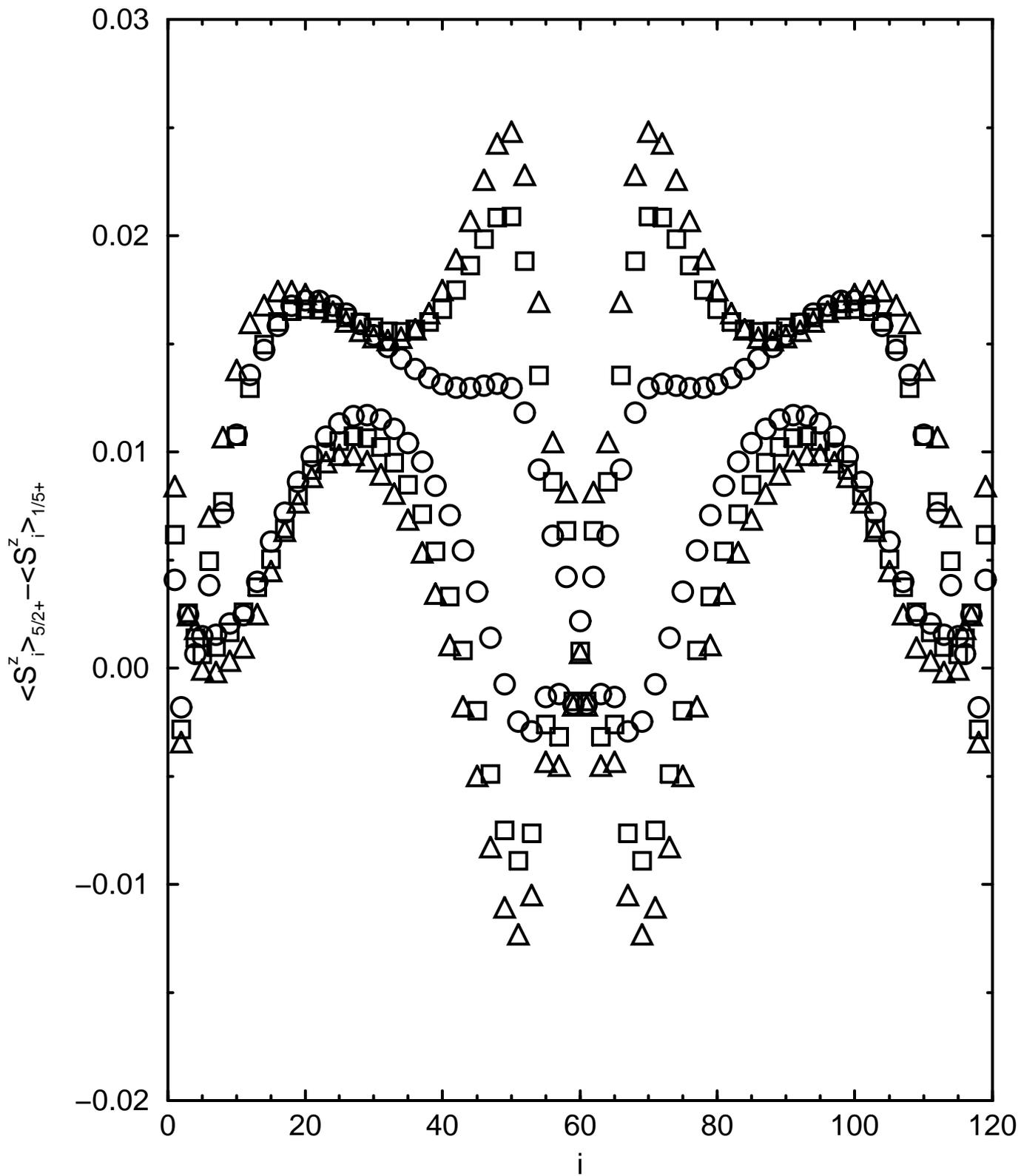
\begin{figure}
\caption{$<S^z_i>_{5/2^+}-<S^z_i>_{1/2^+}$ as a function $i$,
for $J/J'=1$ (circles), $J/J'=0.2$ (squares), $J/J'=0.1$ (triangles).
}
\label{fig:Sz.J1510}
\end{figure}

We have also calculated the expectation value of the z-component
of the spin for the values of $J/J'=1$, $0.2$, and $0.1$.
Our results are shown in Fig.~\ref{fig:Sz.J1510}. The
circles are the results for $J/J'=1$, the squares for $J/J'=0.2$
and the triangles for $J/J'=0.1$. In order to differ between extended
and localized states we have subtracted $<S^z_I>_{3/2^+}$.
As opposed to Section~\ref{sec:bonds} the spin configuration
that we subtract, here $<S^z_I>_{3/2^+}$, is non-zero across
the impurity making it difficult to give a quantitative interpretation
of these results. However, clearly the state is in all
cases extended over the whole length of the chain.

Thus, we conclude that for a strong antiferromagnetic coupling to the
$s=1/2$ impurity no bound states exists below the Haldane gap.
The fact that
DiTusa et al.~\cite{ditusa} observes such states, indicates that, the
appropriate (AF) $J'$, for a simple linear chain Heisenberg model
description of their results, should be rather small.

\section{Ferromagnetic Couplings}\label{sec:ferro.coupling}
We have only performed the numerical DMRG calculations
for antiferromagnetically coupled impurities. It would
be interesting to repeat the calculation using
ferromagnetic couplings.
Due to the rather large amount of CPU time required we
have not done this. Since we in our implementation of the DMRG
method do not work with the total spin as a good quantum number, but only
with it's z-component, ferromagnetic interactions are generally
also more difficult to treat, due to the fact that the ground-state
belongs to a different multiplet for even and odd $L/2$.
However, the effective Hamiltonians
we have obtained in the previous sections can be derived
for a ferromagnetic coupled impurity in a straightforward
way. Below we briefly discuss what one finds.

{}For a central bond that is {\it weak}, $|J'|\ll |J|$, and {\it ferromagnetic}
in an otherwise antiferromagnetically coupled $s=1$ chain,
we expect the results
of Sec.~\ref{sec:weak.link} to still apply. In particular
Eq.~(\ref{eq:deweak}) should still be correct. However,
in this case the ground
state of Eq.~(\ref{eq:heff.weak}) is the triplet.

{}For a central bond that is {\it strong}, $|J'|\gg |J|$,
and {\it ferromagnetic},
the two central spins, ${\bf S}_{L/2}$
and ${\bf S}_{L/2+1}$, will have a $s=2$ ground-state.
We then obtain an
effective Hamiltonian in terms of an effective $s=2$ impurity:
\begin{equation}
H_{\rm eff}=J\sum_{i=1}^{L/2-2}{\bf
S}_i\cdot {\bf S}_{i+1} +J\sum_{i=L/2+2}^{L-1}{\bf S}_i\cdot {\bf
S}_{i+1}+H_{\rm imp}(s=2),
\label{eq:heff5}
\end{equation}
where the term describing the $s=2$ impurity is given by:
\begin{eqnarray}
H_{\rm imp}(s=2)
&=&J[{\bf S}_{L/2-1}\cdot{\cal P}^{(2)}{\bf S}_{L/2}{\cal P}^{(2)}+
{\cal P}^{(2)}{\bf S}_{L/2+1}{\cal P}^{(2)}\cdot{\bf
S}_{L/2+2}]\nonumber\\
&=&\gamma J[{\bf S}_{L/2-1}\cdot{\bf S}^{\prime\prime}_{s=2}+
{\bf S}^{\prime\prime}_{s=2}\cdot{\bf S}_{L/2+2}].
\label{eq:heffs2}
\end{eqnarray}
Here $\gamma$ is
determined by the projection of ${\bf S}_{L/2}$ and ${\bf S}_{L/2+1}$
onto the $s=2$ subspace, ${\cal P}^{(2)}{\bf S}_{L/2}{\cal P}^{(2)}=\gamma
{\bf S}^{\prime\prime}_{s=2}$. It is easy to see that $\gamma=1/2$, thus,
the effective coupling to the $s=2$ impurity is $J/2$.
If the coupling to the $s=2$ impurity is regarded as a free parameter
we would expect bound states to appear below a threshold, $K_c^{(2)}$.
However, in the absence of any numerical data,
we have no way of telling whether $J/2$ is above or
below such a  threshold and we cannot draw any
conclusions. Let us briefly consider the case where the coupling
to the $s=2$ impurity is assumed to be a free variable $K$.
{}For $K\ll K_c^{(2)}$ we can again
use the picture of effective $s=1/2$ chain-end spins to write an effective
low-energy Hamiltonian: $H_{\rm eff}=\alpha K [{\bf S}^{\prime\prime}_{L/2-1}
\cdot{\bf S}^{\prime\prime}_{s=2}+ {\bf S}^{\prime\prime}_{s=2}
\cdot{\bf S}^{\prime\prime}_{L/2+2}]$. As in previous sections, ${\bf
S}^{\prime\prime}_{L/2-1}$ and ${\bf S}^{\prime\prime}_{L/2+2}$ are
now effective $s=1/2$ spins. This low-energy effective Hamiltonian
has the following spectrum: A ground-state triplet with energy $E=-\alpha K3$,
followed by a quintuplet at $E=-\alpha K$, another quintuplet at $E=0$,
and finally a septuplet at $E=\alpha K 2$. For sufficiently small $K$,
$K\ll K_c^{(2)}$, we then
expect bound states in the Haldane gap to occur at energies, $2K\alpha$,
$3K\alpha$, and $5K\alpha$. The effective Hamiltonian, Eq.~(\ref{eq:heff5}),
corresponds to $K=J/2$, since this is not a small coupling,
a low-energy Hamiltonian based
on the $s=1/2$ chain-end spins is presumably not
valid. Since we do not know $K_c^{(2)}$, it is not possible to tell if bound
states occur.

If a $s=1/2$ impurity is coupled with a {\it weak ferromagnetic} bond
to the otherwise antiferromagnetically coupled $s=1$ chain, we expect
bound states to occur. As already mentioned in Sec.~\ref{sec:shalf.weak},
we expect these to have energies
$\Delta E_{1/2^-} = \alpha J'/2,\ \ \ \Delta E_{1/2^+}=\alpha J'3/2$,
with $\alpha \simeq 1.064$.

{}For a $s=1/2$ impurity coupled with a {\it strong ferromagnetic} bond
to an otherwise uniform $s=1$ chain,
we know from Sec.~\ref{sec:shalf.strong} that,
the ground state of $H_{\rm imp}$, Eq.~(\ref{eq:h0.shalf})
should be a sextuplet, $S^P=5/2^+$.
If we by ${\cal P}^{(5/2)}$ denote projection onto the
$5/2$ subspace, we can define a constant, $\delta$, by
${\cal P}^{(5/2)}{\bf S}_{L/2}{\cal P}^{(5/2)}=\delta{\bf
S}^{\prime\prime}_{s=5/2}$. We can then write an effective Hamiltonian
in terms of an effective $s=5/2$ impurity
in the following way:
\begin{equation}
H_{\rm{eff}}=J\sum_{i=1}^{L/2-2}{\bf
S}_i\cdot {\bf S}_{i+1} +J\sum_{i=L/2+2}^{L-1}{\bf S}_i\cdot {\bf
S}_{i+1}+H_{\rm imp}(s=5/2),
\label{eq:heff6}
\end{equation}
where the term describing the $s=5/2$ impurity is given by:
\begin{eqnarray}
H_{\rm imp}(s=5/2)
&=&J[{\bf S}_{L/2-1}\cdot{\cal P}^{(5/2)}{\bf S}_{L/2}{\cal P}^{(5/2)}+
{\cal P}^{(5/2)}{\bf S}_{L/2+1}{\cal P}^{(5/2)}\cdot{\bf
S}_{L/2+2}]\nonumber\\
&=&\delta J[{\bf S}_{L/2-1}\cdot{\bf S}^{\prime\prime}_{s=5/2}+
{\bf S}^{\prime\prime}_{s=5/2}\cdot{\bf S}_{L/2+2}].
\label{eq:heff5.2}
\end{eqnarray}
In this case $\delta=2/5$. The effective coupling to the $s=5/2$ impurity is
then $2J/5$. If we regard this coupling as a free parameter, $K$, we expect
bound states to occur for $K$ below a threshold, $K_c^{(5/2)}$.
We are not able to
determine if $2J/5$ is below such a threshold.
As above, let us consider
the case where the coupling
to the $s=5/2$ impurity is a free variable $K$.
{}For $K\ll K_c^{(5/2)}$ we can again
use the picture of effective $s=1/2$ chain-end spins to write an effective
low-energy Hamiltonian: $H_{\rm eff}=\alpha K [{\bf S}^{\prime\prime}_{L/2-1}
\cdot{\bf S}^{\prime\prime}_{s=5/2}+ {\bf S}^{\prime\prime}_{s=5/2}
\cdot{\bf S}^{\prime\prime}_{L/2+2}]$. As above, ${\bf
S}^{\prime\prime}_{L/2-1}$ and ${\bf S}^{\prime\prime}_{L/2+2}$ are
now effective $s=1/2$ spins. This low-energy effective Hamiltonian
has the following spectrum: A ground-state quadruplet with energy
$E=-\alpha K7/2$,
followed by a sextuplet at $E=-\alpha K$, another sextuplet at $E=0$,
and finally an octuplet at $E=\alpha K 5/2$. For
sufficeintly small $K$ we then
expect bound states in the Haldane gap to occur at energies, $\alpha K 5/2$,
$\alpha K 7/2$, and $\alpha K 6$. The effective Hamiltonian,
Eq.~(\ref{eq:heff6}), corresponds to $K=2J/5$, this coupling
cannot be regarded as small
and we are not in the regime $K\ll K_c^{(5/2)}$.
Thus, the description
based on the $s=1/2$ chain-end spins is of limited use.
However, whether $2J/5$ is below or above $K_c^{(5/2)}$ is an open question.

We see that only in specific cases is it straight forward to derive a useful
{\it low-energy}
effective Hamiltonian. For a strong ferromagnetic bond, and for a
$s=1/2$ impurity coupled with a {\it strong ferromagnetic} bond,
further numerical work will be needed in order to determine
whether bound states in the Haldane gap exists.

\section{Discussion}
We have shown that, by the introduction of an AF bond with
$J'\ne J, J'\ge 0$ in an otherwise isotropic AF $s=1$ chain, a
triplet bound state below the Haldane gap occurs for sufficiently large
and small $J'$. The energy of the bound state is in very good
agreement with perturbative results based on effective $s=1/2$
chain-end excitations created by broken valence bonds. In a region
$0.7J\approx J'_{c\ \rm weak} \le J'\le J'_{c\ \rm strong}\approx 2.0 J$
no bound state is found to within numerical precision.
If a $s=1/2$ impurity is introduced with
an AF coupling $J'\ne J, J'\ge 0$ in an isotropic AF $s=1$ chain we
find that, only for $J'$ below $J'_{c\ 1}\approx 0.45J$ do
bound states occur. For $J'<J'_{c\ 1} \approx 0.45J$ a
bound state appears, and for $J'<J'_{c\ 2}\approx 0.35J$
a second bound state is stable.
At large impurity couplings, $J'>J$, we find no bound states.

In experiments on Y$_{2-x}$Ca$_x$BaNiO$_5$~\cite{ditusa},
states below the Haldane gap have been found. In a simple
linear chain spin model of Y$_{2-x}$Ca$_x$BaNiO$_5$,
our results would imply
that the coupling to a $s=1/2$ impurity, introduced by the Ca doping,
should be rather small, if it is antiferromagnetic.
This may seem surprising since we would expect $|J'|>|J|$
to be the more likely scenario in Y$_{2-x}$Ca$_x$BaNiO$_5$.
However, the coupling could be large and {\it ferromagnetic}
in which case we cannot rule out the existence of bound states.
Secondly, the non-zero single ion anisotropy term, $D(S^z_i)^2$,
relevant for Y$_{2-x}$Ca$_x$BaNiO$_5$ could create bound states
in the gap even for a large antiferromagnetic coupling to the
$s=1/2$ impurity. Also, the substitution of Y$^{3+}$ by
Ca$^{2+}$ will lower the crystal field symmetry on close by
Ni sites, justifying new {\it local} anisotropy terms in a
spin Hamiltonian. Possibly such terms could create bound states.
DiTusa et al.~\cite{ditusa} argue that the magnetic disturbance
caused by the Ca doping introduces either a triplet or a quartet
of subgap bound states. Our results indicate that
several different multiplets of bound states are likely
to be present, some of which have spin larger than $1/2$,
even for the simple scenario of Ca doping giving rise
to $s=1/2$ impurities, thus justifying the large contribution
to the spectral weight from the impurities seen in the experiments.

Our finding that a critical coupling is needed to form
bound states in this problem may seem surprising at first,
since one would expect that even an infinitely small
impurity potential would form a bound state in one
dimension. However, in a NL$\sigma$ model description one
can argue that the introduction of impurities do not
give rise to simple local potentials due to the constraint
$|\bbox{\phi}|^2=1$. This might change in a model where
the rotational symmetry is broken by the introduction
of a single ion anisotropy term, $D(S^z_i)^2$. Such a
term is relevant for the description of NENP and Y$_2$BaNiO$_5$.
It would therefore be very interesting to repeat the above
DMRG calculations with $D\ne 0$.

\acknowledgments
We gratefully acknowledge many enlightning discussions with
W.~J.~L.~Buyers.
One of us (ESS) would like to thank C.~Broholm, D.~Reich,
R.~Hyman, S.~M.~Girvin and  A.~Macdonald for useful
discussions and comments.
ESS is supported by NSF grant number NSF DMR-9416906.
This research was supported in part by NSERC of Canada.

\newpage

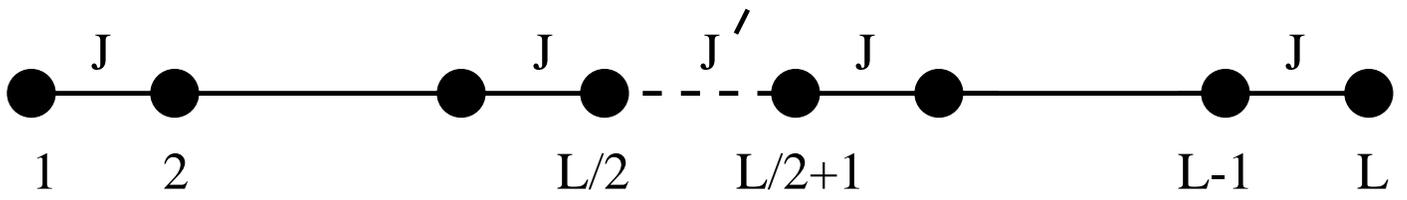
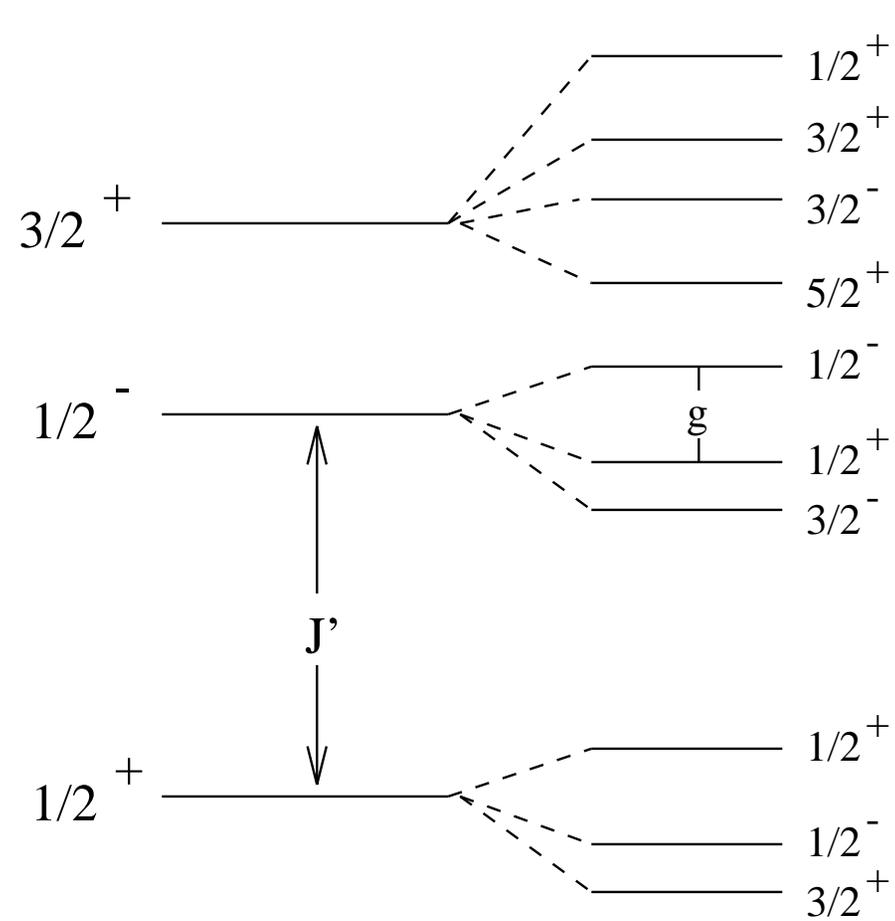
\end{document}